# Is controlling a brain-computer interface just a matter of presence of mind? The limits of cognitive-motor dissociation.

**Perrine Seguin**[1,*], **Emmanuel Maby**[1], **Fabien Perrin**[2], **Alessandro Farnè**[3], **Jérémie Mattout**[1*]

[1] Université Claude Bernard Lyon 1, CNRS, INSERM, Centre de Recherche en Neurosciences de Lyon CRNL U1028 UMR5292, COPHY, F-69500, Bron, France

[2] Université Claude Bernard Lyon 1, CNRS, INSERM, Centre de Recherche en Neurosciences de Lyon CRNL U1028 UMR5292, CAP, F-69500, Bron, France

[3] Université Claude Bernard Lyon 1, CNRS, INSERM, Centre de Recherche en Neurosciences de Lyon CRNL U1028 UMR5292, IMPACT, F-69500, Bron, France

*Corresponding authors: perrine.seguin@inserm.fr, jeremie.mattout@inserm.fr

## Abstract

Brain-computer interfaces (BCI) are presented as a solution for people with global paralysis, also known as locked-in syndrome (LIS). The targeted population includes the most severe patients, with no residual eye movements, who cannot use any communication device (Complete LIS). However, BCI reliability is low precisely in these cases, technical pitfalls being considered responsible so far. Here, we propose to consider also that global paralysis could have an impact on cognitive functions that are crucial for being able to control a BCI. We review a bundle of arguments about the role of motor structures in cognition. Especially, we uncover that these patients without oculomotor activity often have injuries in more "cognitive" structures such as the frontal eye field or the midbrain, exposing them to cognitive deficits further than canonical LIS population. We develop a hypothesis about the putative role of the motor system in (covert) attention, a capacity which is a prerequisite for most BCI paradigms and which should therefore be both better assessed in patients and considered.

**Keywords:** Cognitive motor dissociation; Complete Locked-in syndrome (CLIS); Amyotrophic Lateral sclerosis (ALS); Brain-computer interfaces; Covert attention; Disorders of consciousness; Embodied cognition; Paralysis

# MAIN TEXT

# 1 Introduction

The role of the body in cognition has long been a central philosophical question. At the time of the development of brain-computer interfaces (BCI), which aim to allow completely paralyzed people to communicate, the question arises again, in a very pragmatic way: can a completely paralyzed person control a BCI? In other words, does the loss of motor control impact cognition? And if so, to what extent?

Medical advances in intensive care help to increase the survival of patients, but it is still difficult to predict the functional recovery of patients. Hence, it confronts clinicians with more patients with complete or near complete paralysis, some unresponsive. This raises the dramatic ethical dilemma as of the relevance of implementing life-sustaining treatments in these patients. Such a crucial decision requires attempting everything in order to obtain their auto-determination. But how to know if a patient is aware in the absence of motor response? Is they able to understand our questions? Will they be able to take and communicate an informed decision? These uncertainties amplify the doubt and suffering in relatives and caregivers who face dramatic choices.

To get around paralysis and detect covert awareness, active neuroimaging paradigms are now proposed whereby patient are asked to covertly follow a command during functional brain imaging. Brain-computer interfaces should be the next step after the detection of consciousness, to restore communication.

In this paper, we emphasize that those patients who are still aware are struggling though to communicate by BCIs. Moreover, we show that the efficacy in the use of BCIs is prevented also in those patients who, despite being still able to communicate, present with some oculomotor impairments.

Hence, the current state-of-the-art leads us to question the role of some motor abilities as a support for cognition. Indeed, compared to patients with "classic" locked-in syndrome that can communicate with their eyes, completely paralyzed patients present two additional and important motor impairments: they have lost the control of their gaze and sometimes of their respiration.

Thus, here we interrogate the neuroscience literature to gain a better understanding of the link between oculomotor control and cognition and to grasp its potential implication for BCI use. It is indeed well-established that eye movements are strongly correlated with spatial attention. We make the parallel between the predictions of the premotor theory of attention (Rizzolatti et al., 1987) and the observations made in patients with global paralysis. The premotor theory of attention proposes that spatial attention results from an activation of the neuronal networks that program movements. In line with this theory, we further raise and critically assess the hypothesis that spatial and temporal selective attention abilities may be altered when there is a global loss of motor control.

Relying on empirical findings and theoretical accounts, we question the functional independence of some motor and cognitive systems. Importantly, their intricacy would highlight the need to assess specific cognitive dimensions in some patients' populations, including exogenous and endogenous attention, but also interoception, peri-personal space, agency, metacognition and volition.

## 1.1 *Clinical conditions leading to a locked-in state*

Albeit with different etiologies, the typical end-users targeted by the BCI field have in common their clinical state that is a total paralysis, acquired after an otherwise normal development of the nervous system. Their injuries stand along the cortico-spinal pathways or in the peripheral nervous system.

The 'classical' locked-in syndrome (LIS) is caused by an injury in the ventral part of the pons, most of the time due to stroke (León-Carrión et al., 2002; Plum & Posner, 1972). The patient is totally paralyzed, except for the vertical dimension of eye movements and blinks, which are spared and allow them to communicate. Some of them can even lose the third and VII cranial nerves that allow these residual movements, and then are non-responsive behaviorally (Plum & Posner, 1972; E. Smith & Delargy, 2005).

This condition can also be encountered at the late stage of amyotrophic lateral sclerosis (ALS), a neurodegenerative disease of the motor neurons. Different forms of this disease exist, some affect more the lower motor neurons (progressive muscular atrophy), cortico-spinal motor neurons (primary lateral sclerosis), brainstem motor neurons (bulbar ALS) or cortico-frontal bulbar motor neurons (pseudobulbar palsy) (Taylor et al., 2016). In all these forms, oculomotor muscles are usually preserved, except in the extremely late stages.

Other possible causes of a non-responsive state are severe brain injuries. Etiologies are mainly severe traumatic brain injuries, strokes or anoxia after a cardiac arrest, all of these inducing diffuse brain injuries. After a comatose state, up to four weeks, these patients can remain in a state with preserved vegetative functions (e.g.: autonomous respiration and opening of the eyes), but no sign of awareness of their environment. Clinically, these patients are considered as suffering from disorders of consciousness (DOC). Amongst these persons, some of them could be conscious, but a combination of impairments (motor, sensory, cognitive ones) can prevent command understanding and/or following.

Finally, a massive and global damage of the peripheral nervous system, as observed in the Guillain-Barre syndrome can also lead to a complete locked-in syndrome, which is most of the time temporary.

To refer to these groups of patients with limbs paralysis, anarthria and facial paralysis, we will mainly use the expression "global paralysis". For patients with an additional complete oculomotor impairment, but presumably conscious, we will use the term "complete locked-in syndrome".

### *1.2  Cognitive motor dissociation: detecting "consciousness" in the absence of behavior*

To detect consciousness in non-responding patients (i.e., potential DOC patients), numerous functional neuroimaging protocols, using electrophysiology or functional MRI, have emerged over the past two decades, searching for neuronal signatures of conscious information processing. The first paradigms were inferring consciousness from perceptual or attentional biomarkers such as the P300 electro-encephalographic (EEG) waveform, sometimes without instructions to the subjects (*see 1.3.2 and 2.4.1 for more information on this evoked response*). There can be up to 30 % of patients presenting this kind of dissociation between behavior and cerebral activity (Aubinet et al., 2022; Fernández-Espejo & Owen, 2013; Ferré et al., 2023; Perez et al., 2021; Schnakers et al., 2022). However, it is difficult to infer consciousness content from the presence of these markers. That's why the up-to-date panel of exploration now also includes some command following protocols, which are a first step toward achieving reportability of the content of consciousness by the patients. These protocols are very similar to BCI ones. BCI paradigms go further in the sense that they enable the classification of single trial responses, online, and can thus provide online feedback to the subject.

These active neuroimaging protocols uncovered the possible existence of a cognitive-motor dissociation: the ability to mentally follow commands, without showing reliable motor response during clinical examination (Claassen et al., 2019; Cruse et al., 2011; Monti et al., 2010; Morlet et al., 2022; Owen et al., 2006; Schiff, 2015; Schnakers et al., 2009). The pioneering protocols used fMRI (Owen et al., 2006), but EEG is now more widely used for practical reasons: relatively inexpensive, bedside monitoring is a precious asset in the face of such a fragile population with fluctuations of awareness. The cognitive motor dissociation can be revealed with current protocols only if there is a preserved basic understanding of language, and at least one preserved sensory pathway (auditory or visual) to get access to instructions. Furthermore, attention is often mandatory, as well as motivation, working memory and planification. However, the lack of behavioral response suggests a motor impairment, hence the "cognitive motor dissociation". Note that there is a gradient of cognitive requirements in these active protocols. Some ask for a simple motor command "move your hand" (Claassen et al., 2019; Cruse et al., 2011), close to what is used behaviorally in clinical settings, because presumably more intuitive. It is then difficult to precisely infer the cognitive state of responders. In contrast, protocols requiring two different tasks like mental navigation versus tennis playing suppose more preserved cognitive functions (eg: working memory). Nevertheless the proportion of patients showing cognitive motor dissociation amongst patients with disorders of consciousness is quite stable across protocols, ranging from 10 to 20 % (Kondziella et al., 2016 ; Schnakers 2022). At the acute phase of brain injury, the presence of cognitive motor dissociation is one of the strongest predictor of a better recovery (Egbebike et al., 2022). Importantly, in some protocols, the type of modulation of brain activity is not specific (Claassen et al., 2019; Egbebike et al., 2022): any voluntary modulation of brain activity can be interpreted as a response to command.

## 1.3 *Restoring communication using BCI: a difficult promise to keep*

### 1. *Communication in cognitive motor dissociation is rare and fluctuating*

Once command following is achieved, the objective is to restore functional communication. Monti et al (Monti et al., 2010) report an attempt to use mental imagery tasks for communication: one patient could answer 6 questions. However, this patient remains an exception: he was the only one to succeed in this task, whereas only 5 out of 54 tested patients could follow the command, in the first place. We face here a dramatic situation: since the discovery of covert awareness in 2006 (Owen et al., 2006), patients can be diagnosed as conscious, but we do not have a reliable solution to restore communication with them.

In parallel, there is a growing field of research with clinical brain-computer interfaces trying to restore communication with persons with almost total paralysis. Some of these patients evolve to a "complete locked-in syndrome" (CLIS) i.e., total absence of motor control, but supposedly conscious with (globally) intact cognitive abilities. BCIs implement a real-time analysis of brain activity, and in this field, for the same reasons already mentioned, EEG is arguably the most appropriate technique today. Real-time analysis results in sending feedback to the patient about their own command, and allows them to adapt their brain activity accordingly.

Most studies tend to show that brain computer interfaces work poorly with the most severely paralyzed patients (Birbaumer, 2006; Kübler & Birbaumer, 2008; Lazarou et al., 2018; Marchetti et al., 2013).

Two studies published in 2017 claimed that a communication was restored with complete locked-in patients (Chaudhary et al., 2017; Guger et al., 2017), but some of the methodological aspects of these studies remain unclear, and their results were debated in the BCI community (Spueler, 2018). One paper was even retracted in 2019 (Editors, 2019).

A longitudinal study showed that a person with ALS could learn to use a BCI at the LIS stage and then maintain some control when becoming CLIS. This study included three subjects with ALS in the LIS state, and one became a CLIS during the 27 months follow up (Okahara et al., 2018). This patient, a 37-year-old woman who had ALS for 6 years, was asked to control a binary SSVEP-BCI by either focusing on a LED, or ignoring it. From month three onwards, the electro-oculogram could not show any difference between the attended and ignored conditions (confirming the CLIS). However, she performed BCI control with 79% accuracy online, with accuracies above the confidence limits in 18 out of the 27 months and in 27 out of 40 sessions. Despite this promising result, when used in a communication mode, the answers were not reliable. Another longitudinal study assessed four different paradigms with a person at an advanced stage of ALS who lost her ability to communicate one year before the beginning of the study. The authors chose first to compare different tasks offline: an auditory oddball paradigm, a left motor imagery (LMI), mental subtraction (MS), and tongue motor imagery (TMI). The more reliable tasks (combination of LMI and MS) were selected to be used online. The patient could control the online non-invasive BCI with 87.5% accuracy, over 20 trials (NB: chance level at 70% for that number of trials). Four months after this single online experiment, her cognitive status deteriorated, with no detectable evoked potentials anymore, which prevented any longitudinal BCI and communication assessment (Han et al., 2019). The authors suggested a link between this neurological deterioration and a pneumonia, but other studies found a degradation of the electrophysiological markers in patients in CLIS. For example, a longitudinal observation in electrocorticography was conducted in a patient with ALS, around the time when he lost all muscle control (Bensch et al., 2014). Whereas the P300 wave was still observable at the time of this dramatic evolution, three months later, it was no longer detectable.

A recent publication with an implanted intra-cortical electrode in the dominant left motor cortex demonstrated both the feasibility and the striking limitations of communication with a CLIS patient at the advanced stage of ALS. This patient was implanted once he was already in a CLIS state with no residual eye movements, as attested by EOG (Chaudhary et al., 2022). During the first stages of the training, it appeared that when the patient was instructed to attempt or imagine hand, tongue or foot movements, no cortical response could be detected. Reliable yes-no responses were finally obtained three months after implantation, thanks to a neurofeedback protocol. Tones with two different frequencies were provided according to the neural activity, using a spike mapping rate, with the necessity to adjust the threshold manually at each session. Over the 356 days of training and testing, he obtained an accuracy of 86.6 % on 5700 trials. During training sessions where his accuracy was above 80%, he could then use a speller with auditory propositions of blocks of letters and then single letters. He proved able to spell one letter per minute, and freely produce intelligible sentences on 44 out of 107 days, which allowed him to express some of his needs and positive feedback to the researchers about the BCI.

To our knowledge, this is the only study with intra-cortical electrodes at the CLIS stage. Intra-cortical electrodes are so far the best that we can expect in terms of signal-to-noise ratio in BCI, and it is then striking that even in this optimal condition, the number of commands had to be reduced such as the BCI was finally binary, with a low information transfer rate, and that performance fluctuated so much.

In conclusion, despite the development of various strategies for enabling BCI control, it is noticeable that no reliable and efficient long-term communication was achieved with neuroimaging so far, which maintains the doubt about the integrity of cognitive abilities of the persons in CLIS. This important dissociation between command following and functional communication is also seen in standard clinical assessments. In the most widely used scale to detect consciousness, the Coma Recovery Scale Revised (CRS-R), the emergence from minimal consciousness is defined by the recovery of functional communication, or by the functional use of a tool. Command following indicates at least a minimal conscious state. A finer analysis by Naccache of the items of CRS-R highlights that response to command is associated with cortical activation, but does not guarantee consciousness. To avoid confusion, he proposed to replace the expression "minimally conscious state" by "cortically mediated state" (Naccache, 2018). This expression refers to patients who present an activation of the cortex, without presuming what is their "state of consciousness". In the face of current evidence, with the lack of sustained and reliable functional communication, we argue that the great majority of persons in a cognitive-motor dissociation state may indeed be viewed as being in a cortically mediated state rather than in a conscious state with functional communication. In the following section, we briefly review the main paradigms of BCI, in order to have a better idea of the cognitive prerequisites and therefore of the deficits which, in patients, could explain the significant gap that remains between the promises made and the results achieved.

## 2. *Controlling a non-invasive BCI: multiple paradigms but a common attentional demand*

EEG suffers from a lack of spatial resolution and covers a restricted frequency span. Command detection thus relies on mental efforts that can either recruit fairly large and distinct cortical networks, or induce distinct dynamics, or both. Moreover, reliable classification results from a trade-off between the amounts of data that have to be acquired to infer a single command, and the time it takes to complete a single decision. More importantly, controlling an active BCIs requires the ability to select and plan a mental action, as well as to orient and sustain a substantial attentional effort. Traditionally, a distinction is made between endogenous and exogenous BCIs for communication. The former are controlled thanks to an attentional modulation of external stimulations (e.g. P300 and steady-state evoked potential BCIs), while the latter are independent from sensorial afference (e.g. mental imagery, mental calculation based BCIs).

*P300:* The most exploited neurophysiological marker in this context is arguably the P300, a positive wave observed in response to external stimuli (about 300ms later), if relevant to the user and relatively rare (among irrelevant or distracting stimuli). If the "P300-speller" mainly uses the visual modality, the P300 wave can also be observed after auditory or tactile stimuli (Allison et al., 2020). For selection, the patient has to actively attend the expected target stimulus, i.e., to selectively pay attention to it. This task can be done overtly (gaze-dependent) or covertly (gaze-independent). Most P300 BCIs are in fact not restricted to measuring the P300. For instance, they typically also exploit modulations of the N200. Both N200 and P300 are modulated by covert attention (Treder & Blankertz, 2010). Besides, several studies showed that classification results are better when subjects are allowed to direct their gaze toward the target (Brunner et al., 2010; Treder & Blankertz, 2010). Numerous visual P300 BCIs have been assessed in patients with global paralysis, but so far these tools are not used in daily life (Allison et al., 2020).

Some tactile BCIs were assessed with persons in LIS, with the hope to shortcut the oculomotor problem, and obtained a control above chance level (Kaufmann et al., 2013; Severens et al., 2014). However, we found no longitudinal report of daily life use.

*Steady-state evoked potential (SSEP)*: SSEP are characterized by a rhythmic neural response to a flickering stimulus at the frequency of the flicker and its harmonics (Regan & Regan, 1989). It is possible to use the visual channel (SSVEP) or the somatosensory one (SSSVEP). SSVEP BCI protocols use several targets with different flickering frequencies, whereas tactile SSEP use different vibrating frequencies. The amplitude of the SSEP is increased when the subject pays attention to the target (Davidson et al., 2020), which allows to detect which stimulus the subject is attending. It appeared that visual SSEP paradigms can be difficult for patients with LIS (Lesenfants et al., 2014). As for SSSEP-BCI, the clinical trials are sparse.

*Motor imagery:* The kinesthetic imagination of a movement produces a brain activity quite similar to the one observed during the actual production of that same movement (Pfurtscheller et al., 2000). A desynchronization can be detected in the EEG, with a decrease in signal strength in the mu (8-12 Hz) and beta (12,5-30 Hz) frequency bands. These signals occur both during the preparation and the execution or the imagination of a movement. It is measured over the primary motor cortex, mainly on the contralateral side of the movement. The spatial resolution of EEG is low, but it is possible to distinguish a desynchronization of the left motor cortex (associated with a right hand movement) from a desynchronization of the right hand cortex (associated with left hand movement). If the number of commands remains limited, many publications have shown, in healthy volunteers, the ability to control a gauge on a screen (Pfurtscheller et al., 2003).

However, these interfaces require a long training even in neurologically preserved conditions, offer limited performance and above all a large number of healthy volunteers (around 30%) would be unable to produce the desired signals, a phenomenon referred to as "BCI illiteracy" (Vidaurre & Blankertz, 2010).

The mu rhythm is not specific of motor control. It is also implied in attentional tasks (Ross et al., 2022), and its power is negatively correlated with the activity of the attention control network (Yin et al., 2016). Note that even for invasive BCI, it appears that motor cortex activity is contaminated by attentional shifts (Gallego et al., 2022).

Detection of sensorimotor rhythms is better when the subjects are allowed to move their head and their eyes (Bibián et al., 2022). This could be explained partly by the fact that EEG is contaminated by changes in EMG activity during attentional effort, especially for frequency above 20 Hz (Whitham et al., 2007, 2008). But it could also be due to the fact that motor activity facilitates the apparition of these rhythms synchronization or desynchronization.

*Slow cortical potentials:* These features are a slow drift of the EEG signal that reflects a change in cortical excitability. A depolarization of a large neuron assembly corresponds to greater excitability, whereas a hyperpolarization corresponds to a greater inhibition. These signals were among the first used in patients. Several healthy subjects and patients managed to communicate using a BCI exploiting these potentials, but at the cost of several months of training and with a writing speed of the order of one character per minute, well below the means for alternative communication when these are accessible (Kuebler et al., 1998; N. Neumann et al., 2003).

*Mental calculation:* Another way to have a self-paced control over a BCI is to perform mental calculation (Bojorges-Valdez et al., 2015), which activates prefrontal, parietal and lower temporo-occipital regions in healthy subjects (Vansteensel et al., 2014). Up to now this task has been rarely explored using EEG in patients with severe motor disability, but has been explored by a few teams in invasive (ECoG) studies as an alternative when patients had failed to control BCI with motor imagery (Leinders et al., 2020).

*Conclusion:*

As highlighted in this section, the signals used as BCI commands all rely on some form of selective and sustained attentional process. Moreover, the ability to move a little the eyes or the head leads to better classification results, be it for P300 or motor imagery paradigms.

At this point, one should argue that, considering the prevalence of cognitive impairments of persons with severe motor disabilities (Beeldman et al., 2016; Conson et al., 2008, 2010; Rousseaux et al., 2009; Schnakers* et al., 2008), it is no wonder that they struggle to control BCI with such demanding paradigms.

However, as we will see in the next section, it is striking to see that a strong predictor of BCI performance seem to be one of their motor abilities: the oculomotor control.

### *1.4  Is BCI control easier with residual motor control? The key role of oculomotricity*

The press often emphasizes publications reporting attempt to restore communication with LIS patients through non-invasive BCI, regardless of the real scope of the scientific results. Truth is that BCI have routinely been used by patients in very rare cases only (Lazarou et al., 2018; Wolpaw et al., 2018). Researchers and clinicians point to the lack of usability of current systems that are still difficult to use on a daily basis without the help of experts (Kübler et al., 2014; Nijboer, 2015).

However, looking closely at the results of the few studies conducted with patients, and despite their heterogeneous designs and conclusions, one important aspect stands out. Indeed, as we will highlight in the following sections, these interfaces can work well, on average, in severely disabled patients provided they can use another type of interface (e.g., a gaze tracking system), thanks to their residual muscle activity. Conversely, demonstrations of restoration of communication with a BCI remain extremely rare and fragile in patients who need these BCI because they have no or almost no means of communication (i.e. with abnormal oculomotor movements or with no oculomotor movements at all (CLIS)).

One of the rare BCI study providing a flow chart of the inclusions of all patients revealed that amongst 37 persons at a late stage of ALS, 24% (n=9) could not control a P300-BCI (Wolpaw et al., 2018). With a motor imagery BCI, a recent study showed that only one LIS patient out of five succeeded in controlling a BCI while attempting to repeatedly execute feet dorsiflexion and still, the achieved accuracy was low (64% over 45 trials) (Lugo et al., 2019). BCI studies involving healthy subjects, LIS and ALS patients uncovered that patients' performances are often lower, in terms of both communication speed and accuracy (Bai et al., 2008; Holz et al., 2015; Maby et al., 2011; Mugler et al., 2010; Nam et al., 2012). This was also observed with persons with spinal cord injuries (Jeon & Shin, 2015), albeit the lack of data to assess whether the effect size is of the same amplitude.

The high prevalence of oculomotor disorders amongst these patients progressively appeared as the main problem for visual P300-BCI, leading in some studies to the exclusion of these patients (McCane et al., 2015). Hence, auditory BCI with spatialized stimulations were proposed for "gaze-independent" control (Lulé et al., 2013; Sellers & Donchin, 2006). Surprisingly, in these few clinical studies, including ours, most patients continue to show poor performance compared to healthy subjects (Séguin et al., 2023). As for current visual but gaze-independent paradigms, they also fail to provide an effective solution to the patients (Lesenfants et al., 2018).

This is confirmed by some visual BCI studies addressing the problem of covert attention. Be it in healthy subjects (Brunner et al., 2010) or patients with oculomotor impairments (Marchetti et al., 2013), it was observed that P300-BCI performance are much higher and involve much less mental workload when the gaze can be directed toward the target.

Taking all these data into account points to the importance of at least partially preserved oculomotricity to be able to control a BCI. More precisely, as we will develop in the following, the loss of control of the eyes in total paralysis represent a dramatic impairment of the ability to actively sample the environment, even covertly.

## 2  The myth of a clear cut between motor and cognitive functions

There is a growing amount of evidence showing the implication in cognitive tasks of structures initially labelled as "motor" or "sensory". This appears to be true for not only for the cortex but also for many subcortical structures. We cannot review here all the existing literature about the so-called embodied cognition (Barsalou, 2008), but see for more information (Janacsek et al., 2022; Mendoza & Merchant, 2014). We briefly recall here the main "motor structures" that can be directly injured or indirectly disturbed in global paralysis.

### 2.1  *Functional anatomy*
#### *1. Motor and pre-motor cortices*

The motor cortex is divided in sub-parts, some of them more dedicated to motor planification, preparation or execution. These main parts are the primary motor cortex, the premotor cortex and the supplementary motor area. Albeit initially discovered for their implication in the motor control, it happens that all these regions can also be activated during varied cognitive tasks, including overt and covert attention (Avenanti et al., 2012; Bonnard et al., 2004; Freedman & Ibos, 2018; Gallego et al., 2022; Kukleta et al., 2016; Orgs et al., 2016; Vb & Rj, 2019). This entanglement between motricity and locus of attention is increasingly referred to as one of the cause of classification problem when attempting to decode movement intention in primary motor cortex, even with invasive technics (Freudenburg et al., 2019; Gallego et al., 2022; Kaufman et al., 2014; Lebedev, 2017).

Motor networks are also strongly implicated in temporal perception and temporal attention (De Kock, Gladhill, et al., 2021; Merchant et al., 2013). A recent methodological advance studying the edges of brain networks uncover that the sensorimotor and attentional networks presented the greatest level of overlapping clusters (Faskowitz et al., 2020).

The upper motor neurons located in the primary motor cortex are degenerated in ALS. In the "classical" LIS, they also degenerate due to an injury of their axon at the level of the brainstem.

### 2. Frontal eye-field

The frontal eye field is considered as a "crossroad for eye movements and visuo-spatial cognition" (Vernet et al., 2014). Lesions in the FEF mainly alter eye movements for which a voluntary component is introduced. The visual activity encoded within the FEF seems to be related to the computation of a saliency map (Thompson et al., 1997; Walker et al., 2009). FEF seems to be an important component for coupling the voluntary orientation of attention and the oculomotor movement. Beyond overt attention, its role in covert attention was also uncovered in several studies (Armstrong et al., 2009; Kodaka et al., 1997; Monosov et al., 2008; Thompson, 2005 ; Ro et al, 2003).

The frontal lobe, and hence the FEF, are injured in advanced stages of ALS (Reyes-Leiva et al., 2022). On the contrary, it is normally preserved in classical LIS.

### 3. Basal ganglia and Cerebellum

Basal ganglia and cerebellum are crucial for the motor control, but there is also a growing amount of evidence of their important implication also in cognitive processes (Krauzlis et al., 2018). All cortical areas project on the basal ganglia, and these afferents all originate from the cortical layer 5. This convergence places the basal ganglia as both a bottleneck and a highly integrative structure for attention and motor control (Sherman & Usrey, 2021).

Basal ganglia and cerebellum are also densely interconnected between themselves, and form an integrated network with the cerebral cortex, and it is shown that a disorganization of one node have repercussion on the others (for a review, see (Sherman & Usrey, 2021)).

Basal ganglia are atrophic in ALS (Bede et al., 2013). Focal lesions of the cerebellum are observed in ALS, but it is difficult to know if they are secondary do motor pathways re-organizations, or a degeneration due to genetic factors (Bede et al., 2021). Ponto-cerebellar fibers are directly altered in the classical LIS (Leonard et al., 2019).

### 4. Superior colliculus

The superior colliculus is part of the midbrain, and a sensorimotor structure, responsible for the mapping between visual input and the transformation in motor commands. It sends the oculomotor command directly to the oculomotor nuclei in the brainstem. It is also responsible of a more general orienting response of the head and the upper limbs (Allen et al., 2021). Beyond this role, it is also involved in covert attention (Ignashchenkova et al., 2004; Kustov & Robinson, 1995), and in decision making (Basso et al., 2021; Basso & Wurtz, 1997; McPeek & Keller, 2004). The superior colliculus receive input from most of cortical regions and from the basal ganglia. It projects on the thalamus and on basal ganglia.

The superior colliculus is not directly injured in ALS, but some authors suspect that its functioning could be indirectly altered due to the severe modification its networks (basal ganglia and frontal cortex notably) (Becker et al., 2019). It is not either directly impaired in classical LIS, where it is mostly its output that are altered (oculomotor nuclei). We did not find any information on possible changes of this structure in case of oculomotor paralysis.

### 5. Thalamus

The thalamus is not supposed to be directly injured in patients with severe motor disability. However, after an injury of the motor cortex, a thalamo-cortical dysrhythmia appears (Wijngaarden et al., 2016). It is unclear if this dysrhythmia is also present in cortico-spinal injuries in the pons, but changes in brain oscillations are also observed in the locked-in syndrome and in ALS.

This disturbance in the oscillation of the thalamus could impact its function in attention.

### 6. Gain control of non-motor cortical regions by subcortical sensori-motor centers

The cortex is widely modulated by automatic activities such as breathing or walking (Dipoppa et al., 2018; Erisken et al., 2014; Herrero et al., 2018; Vinck et al., 2015). It is particularly striking to see the differential impact between nasal and mouth breathing on local field potentials in rodent and human (Herrero et al., 2018).

The impact on cognition of such phenomenon is debated in healthy subjects. Conversely, in global paralysis, the respiratory function is often impaired, which has an impact on cognition (Schembri et al., 2017). This could be due to the mental workload induced by dyspnea, to the lack of oxygen, or to the sleep impairment due to sleep apnea. But additionally, for patients with no nasal airflow due to a tracheotomy, there is probably a change in these slow cortical modulations that has to be explored.

### 7. Evidence for interferences between motor and cognitive processes

We highlighted that most of "motor" regions are activated in cognitive tasks. How does this activation interfere with motor activity? We expose here some results obtained in the field of research on cognitive motor interference. Cognitive-motor dual tasks can lead to a deterioration of performance in each task, be it in patients (McIsaac et al., 2018) or healthy subjects (Yogev-Seligmann et al., 2008). Examples of such tasks are to participate in a conversation while walking, which forces the subject to slow down or prevent him from correctly avoiding obstacles. These paradigms can be used in research studies, as a mean to explore the functional relevance of some sensory-motor processes in higher level cognition (Ostarek & Bottini, 2021).

So far, no specific neural correlates of cognitive-motor interference could be clearly isolated, as additional areas activated in different studies appear to be highly task dependent. This speaks in favor of a systematic implication of cognitive networks, especially attentional and executive function ones, during motor tasks (McIsaac et al., 2018). In other terms, motor processes have an impact on cognition, and vice versa. We observe that optimization of higher-level cognitive activities can require the motor actions to be "congruent" with them. This congruence is often not an immobilization, but a pacing to follow the "rhythms" or the direction of the cognitive tasks. For example, mental spatial navigation is accompanied by specific eyes movements (Fourtassi et al., 2017). Counting is associated with an increased cortico-spinal excitability of hand muscles (Andres et al., 2012; Sato et al., 2007).

In LIS, the residual motor pathways composed by the control of the eyes and eyelids could partly engage in this coupling between motricity and cognition. But the case of the CLIS is a very particular case for research on embodiment, as there are no possible compensation strategies or they are at least very much reduced. The ability to produce motor actions "congruent" with cognition is lost in these patients, even for most of micro-movements.

In sum, in healthy subjects, even when a motor output is not required, the motor structures in the brain are recruited in many different (covert) tasks. But most of the studies leading to these observations are observational, and not interventional studies with selective inactivation of each region. Hence, the precise role of these structures still has to be determined. Of note, we do not know either if the neuronal subpopulations implied in cognitive tasks in these regions are the same that contribute to motor tasks. Moreover, the plasticity other these networks in global paralysis remain largely unknown. As no motor activity is possible, how do these regions evolve? Is the preserved part of this networks still activated during covert cognition?

## 2.2 *Motor activity modulates perception*

Nowadays, most prevalent theories in neurosciences consider perception as an active phenomenon (Schroeder et al., 2010; Yang et al., 2016): sensors are directed or tuned toward the most task-relevant source of information. Active sensing is supported by motor and attentional sampling "routines". At the neurophysiological level, this results in an adjustment of brain rhythms in sensory cortices (Schroeder et al., 2010). These frameworks highlight the strong entanglement of the processes underlying perception, attention, cognition and action. We will discuss here how total paralysis could impact these processes.

### 1. Peripersonal space: The ability to act on the environment changes perception and attention

The peripersonal space was first defined as the space closely surrounding the body. It happened that this space is influenced by many factors, which led some authors to reconceptualize it as « a set of graded fields describing behavioral relevance of actions aiming to create or avoid contact between objects and the body" (Bufacchi & Iannetti, 2018). Peripersonal space is a core concept to understand the link between perception, action selection and motor control: stimuli are processed differently and faster if they are close enough, we can act on them. There is an improvement in accuracy and shorter reaction times to sensory stimuli located within the peripersonal space versus outside (Blini et al., 2018; Zanini et al., 2021). Interestingly, there is a stronger defense response (i.e., eye blink) for nociceptive stimulations applied to the hand, when the hand is placed close to the face compared to when it is far from it (Sambo et al., 2012). Peripersonal space is mainly encoded in brain regions that prepare actions, as the premotor cortex, the putamen and the parietal cortex (Avenanti et al., 2012; Brozzoli et al., 2014; Makin et al., 2012). Immobilization of the arm of healthy subjects for 10 hours induces a reduction of peripersonal space representation in the vicinity of this arm (Bassolino et al., 2015). Studies assessing peripersonal space in patients with severe motor disability (Avenanti et al., 2012; Bartolo et al., 2014) or with patients with disorders of consciousness (Noel et al., 2019) are just in their infancy. The rationale behind these studies is in the same line of our hypothesis: changes in motor abilities could change the perception of stimuli situated inside the peripersonal space, and more generally of stimuli associated to the preparation of an action.

### 2. Sense of agency without moving?

To succeed in a selective spatial attention task, one needs a representation of oneself, i.e. of one's own body and its state, combined with a representation of the environment. A sense of agency is also required: the subject has to experience oneself as the cause of the orientation of attention. Motricity is suspected to play a great developmental role in the construction of self and of agency (Di Paolo, 2019; Haggard, 2017), but once these abilities are developed, can we fully sustain them in the absence of voluntary motor control? There are some studies of agency in hemispheric stroke that tend to indicate abnormal agency in these patients (Miyawaki et al., 2020b, 2020a), but it is difficult to disentangle specifically the role of cortico-spinal tract from other parts of the cortical network.

### 3. Coupling between brain rhythms and vegetative rhythms subserves perception, attention and action (and can be altered in case of severe motor paralysis)

Patients with global motor disability often undergo a disorganization of autonomic functions, from digestion, to bladder management and cardio-vascular systems (Hou & Rabchevsky, 2014). Patients with spinal cord injuries, as in tetraplegia for instance, report an altered perception of their inner sensations, i.e. modified interoception (Scandola et al., 2020). Autonomic dysreflexia that are present in some of these patients could further impact the quality of their interoception. Interoception is rarely addressed in clinics but appears to be linked to the perception of the environment, and of one's own body (Azzalini et al., 2019, 2019; Tsakiris et al., 2011; Zamariola et al., 2017).

Moreover, it appears that people tend to modulate their breathing, during both overt and covert action, such as during mental imagery (Decety et al., 1993; Park et al., 2022) or when listening to music (Bernardi et al., 2006). They tend to synchronize the breathing with the rhythm of the music (Haas et al., 1986). This synchronization with external stimulations tends to improve performance in visuo-spatial tasks (Kluger et al., 2021; Perl et al., 2019). Breathing is under both voluntary and autonomous control. Voluntary control of breathing is mostly mediated by motor structures: primary sensory and motor cortices, the supplementary motor area, cerebellum, thalamus, caudate nucleus, and globus pallidum and the medulla (Criscuolo et al., 2022; McKay et al., 2003). This voluntary modulation of breathing also has an impact on the heart rate variability, as this one is modulated by inspiration. This finely tuned coupling state between external stimulations, brain and body rhythms is not accessible to paralyzed patients with a complete ventilatory assistance. This could impact perception of both themselves and their environment.

## 2.3 Clinical and other empirical evidence of the difficult disentanglement between motricity and cognition

### 1. Persons with severe motor disabilities often present with sensory and cognitive impairments

#### 2.3.1.1 Being in CLIS pre-suppose injuries in more "cognitive" structures

We summarize in *Table 1* the respective injured structures in different pathologies were global paralysis is in the foreground.

| Etiology | Causes | Main/typical localisation of the injuries | Secondary localisation of injuries | Main oculomotor impairment | Putative mechanisms of complete oculomotor paralysis |
|---|---|---|---|---|---|
| **Amyotrophic lateral sclérosis** | Degenerative | Upper and lower motoneurons | Fronto-temporal lobe | **Early stages:** Changes in saccadic intrusion (enlarged) | **Extremely advanced stages:** advanced frontal degeneration (especially frontal eye field) or degeneration of oculomotor nucleus |
| **"Classical" locked-in syndrome** | Vascular, tumor, traumatic, infectious | Brainstem: ventral pons. May extend to medulla or midbrain<br>– Cortico-nuclear tracts<br>– Corticospinal tracts<br>– Cranial nerve VI fascicles<br>– Pontine nuclei and pontocerebellar fibers<br>– Paramedial pontine reticular formation | Depending on the etiology (vascular injuries in other part of the brain for example) | **Classically, mainly at the acute stage:** lateral gaze paralysis (VI) leading to diplopia<br><br>**Additionnal impairments:** oscillopsia, abnormal pupillary light response<br><br>(Graber, 2016) | Additional rostral injury of the nucleus of IIIrd cranial nerve (loss of blink and vertical gaze) in the midbrain. |
| **Guillain-Barré** | Dysimmune, often happening after a viral infection | Myelin +/- axons of nerves and roots | **Miller-Fisher syndrome:** lower cranial and facial nerves<br><br>**Bickerstaff syndrome:** brainstem encephalitis | Prevalence of 15%, mainly associated with Miller-fisher or Bickerstaff syndrome. | Complete demyelination +/- axonal injury of oculomotor nerves in the pons and midbbrain. |

*Table 1: Summary of injuries in different etiologies leading to a global paralysis*

In case of an oculomotor paralysis, there is a higher prevalence of injury in more cognitive parts of the brain, especially for ALS. Indeed, in this disease, the frontal impairment appears early, and is likely to impair oculomotricity before a direct injury on oculomotor nerves, which occur only at the very late stages of ALS, when patients decide to undergo a mechanical ventilation.

Concerning locked-in syndrome, oculomotor impairments are more likely due to a direct injury of oculomotor nucleus in the brainstem.

### 2.3.1.2 Being in LIS is associated with some sensory and cognitive impairments

| Etiology | Cognitive impairments? | Somatosensory impairment? | Visual impairments? | Auditory impairments? |
|---|---|---|---|---|
| **Amyotrophic lateral sclérosis** | Variable, from normal, to fronto-temporal dementia See (Kellmeyer et al., 2018) for a recent review more focused on advanced stages and electrophysiology. | Subclinical, can be detected on evoked potentials (Iglesias et al., 2015; Sangari et al., 2016) | Subclinical, most of visual tests reveal similar results as controls (Moss et al., 2016), but there are changes in visual evoked potentials (Münte et al., 1998) and in fMRI visual cortices activity (Lulé et al., 2010) | Subclinical, changes in evoked potentials (Pekkonen et al., 2004; Raggi et al., 2008) and fMRI activity of secondary auditory areas (Lulé et al., 2010). |
| **"Classical" locked-in syndrome** | Variable, frequent impairment in at least one subscale (Rousseaux et al., 2009; E. Smith & Delargy, 2005) | Possible in case of extension of the injury toward sensory pathways (more posterior in the brainstem). Can be detected on somatosensory evoked potentials. (Bassetti et al., 1994) | Most of the patients with LIS have a visual impairment (Graber et al., 2016) | Variable, can be detected both at clinical assessment (Rousseaux et al., 2009) and in evoked potentials (Verhagen et al., 1986) |
| **Guillain-Barré** | Variable, with possible alteration of evoked potentials at the acute stage (Ragazzoni et al., 2000). Retrospective feedbacks of patients about hallucinations (Forsberg et al., 2008) or amnesia (Ragazzoni et al., 2000) | Highly prevalent. | Occurs only if there is an impairment in oculomotricity. | Variable, can be abnormal, especially in most severe state (non-responsive) (Ragazzoni et al., 2000) |

***Table 2:*** *Summary of cognitive and sensory impairments in different etiologies leading to a global paralysis*

Neuro-ophthalmological evaluations suggest visual impairment in all subjects with classical LIS (Graber et al., 2016). This is a multifactorial impairment including binocular diplopia or oscillopsia, refractive errors, dry eye syndrome, keratitis or visual field defect. Some of these troubles seriously prevent the use of visual BCI.

Tactile perception is usually spared in LIS patients, as the typical cases are due to bilateral ventral pontine lesions, which usually preserve the posterior part of the brainstem that mediates sensory inputs (Plum & Posner, 1972). However, we have few information about somatosensory evoked potentials in this population. As for audition, auditory recognition and oral comprehension of complex sentences were found to be significantly impaired in one study with 9 persons in LIS (Rousseaux et al., 2009).

In ALS patients, the oculomotor system was classically considered as spared (Brownell et al., 1970; Hudson, 1981), but the development of life support systems, exceeding the natural course of the disease, led to observation that oculomotor impairments can be present at the late stage (Mizutani et al., 1992).

These impairments can be both peripheral (nuclear) and supranuclear. It is noticeable that the oculomotor system is crucial to maintain interaction with the world. An earlier onset of oculomotor troubles in the ALS has a negative pronostic valence: it is correlated with a higher probability to progress from the locked-in state to CLIS (Hayashi et al., 2013; Nakayama et al., 2013). If only around 10% of persons with ALS complain about sensory impairment, some studies found up to 23 % present electrophysiological signs of sensory nerve impairment (Iglesias et al., 2015; Isaacs et al., 2007; Pugdahl et al., 2007).

The neuropsychological examination of these patients is limited by the lack of communication. Some adaptations of the usual technics were necessary to overcome this impairment (N. Neumann & Kotchoubey, 2004) and to have a better overview of their neurological status. The dissemination of neuroimaging techniques also helped to refine the patients' cognitive status.

In classical LIS, the review of cognitive impairments reveals heterogeneous cognitive conditions. These ones could be related to different size of brainstem injury, or to the different medical conditions of the subjects which could induce more variability (long stay in intensive care, association with other localization of strokes, etc).

Early studies reported intact cognitive abilities in LIS with classical neuropsychological assessments (Allain et al., 1998; Cappa et al., 1985; Cappa & Vignolo, 1982; Gayraud et al., 2015). One study showed impairments in five out of ten patients that could be related to additional cortical or thalamic structural brain lesions (Schnakers* et al., 2008). However, studies testing embodied cognition hypothesis found some abnormalities. For example, LIS subjects have severe troubles to mentally manipulate hand images (a task that imply a mental manipulation of the subject's own hand), whereas they are still able to mentally manipulate images (Conson et al., 2008). Contrary to healthy subjects, having the presentation of the hand in the correct physiological side (e.g., left hand on the left) doesn't help them to solve the task. This failure of action simulation was interpreted as a defect of embodied cognition. These aspects, namely modifications of the body schema, are usually not tested in these patients.

ALS is not only a motoneuron disease. At the protein level there is a dissemination of TDP43 aggregates (correlated with the ALS progression) first in motor neurons, but also in non-motor ones. Clinically, there is a continuum between ALS and other degenerative disorders, the most frequent being fronto-temporal dementia. About 20% of ALS patients meet the criteria for fronto-temporal dementia. In addition to this group, 30% of ALS patients, without fulfilling the criteria for dementia, showed deficits in fluency, language, social cognition, executive functions and verbal memory. Amongst all the function tested, only visual perceptive functions were preserved (Beeldman et al., 2016). Importantly though, selective attention is already impaired at early and moderate stages of the disease (Mannarelli et al., 2014; Vieregge et al., 1999; Volpato et al., 2016). This is usually attributed to a concomitant alteration of frontal lobe function. However, the prevalent models of the spreading of ALS present the degeneration of the motor cortex and of the brainstem and spinal cord as the first step of the disease. Hence a direct participation of motor and premotor cortex in attention would be compatible with these findings.

It is common to consider that peripheral injuries do not lead to cognitive impairments. The truth is that if there is a massive and brutal alteration of peripheral nervous system, as in Guillain-Barré, this can lead to disorders of awareness, associated to hallucinations.

Of courses, in all these pathologies, they are drastic changes in crucial autonomous functions such as respirations, blood pressure, and numerous medications with side effects. All of these changes could have an impact on cognition. The impact of massive motor neurological changes per se on cognition still has to be explored.

### 2.3.1.3 Neuroimaging and electrophysiological data: functional brain changes in global paralysis

Neuro-imaging obtained in LIS reveal some abnormalities that are not yet well explained. Cortical neuronal synchronization mechanisms in the resting state condition are altered (Babiloni et al., 2010). In a passive P300 paradigm with own name, in 2006, Perrin et al reported that one person in LIS out of four did not present a P300, which was attributed to a poor signal quality (Perrin et al., 2006).

Lugo et al, 2016, reported an absence of P300 in four persons with LIS subjects amongst seven tested in a passive condition where subjects were only asked to listen only. In an active condition, where the subjects were asked to count the deviants, there were still two LIS subjects out of seven who did not present a P300 (Lugo et al., 2016). Finally, an anatomical MRI study showed a selective cortical volume loss in a group of 9 patients with LIS compared to 44 healthy subjects, in regions that could be linked to an alteration of the mirror neuron system (Pistoia et al., 2016), further supporting the idea that embodied process is altered in these patients.

Persons with ALS show alterations of the long range connectivity, as well as local alteration, with decreased connectivity in sensorimotor, basal ganglia, frontal and parietal areas (Basaia et al., 2020). A recent MEG study showed that the late stage of ALS is associated with higher connectivity in all frequency bands, more scale-free and disassortative brain networks (Sorrentino et al., 2018). The event-related potentials are different compared to control subjects, namely the location and amplitude of the late positivity, the amplitude of the early negativity (N200), the latency of the late negativity (McCane et al., 2015) and the mismatch negativity (related to automatic change detection) (McMackin et al., 2019) (for a recent review of patients at an advanced stage, see (Kellmeyer et al., 2018).

### 2. Perceptual and cognitive consequences of impaired movement execution

Owing to the above gathered evidence, it is legitimate to ask whether patients' cognitive or perceptual impairments are solely due to other factors than motor injury. That's why it is important to assess the potential cognitive consequences coming with the impossibility to execute voluntary movements in the absence of injury to the central motor system, for example in the case of temporary (plaster) or permanent immobilization from peripheral lesions (e.g., nerve injury).

#### 2.3.2.1 Impaired motor execution alters perception

Vision is impossible in the absence of motion. Our feeling that we can fixate our eyes proved to be wrong: there are fixational eye movements that we cannot prevent: microsaccades, ocular drifts, and ocular microtremor (Klein & Ettinger, 2019). When the image is artificially stabilized on the retina, removing the effect of saccades and fixational eyes movements, the observer becomes blind. This effect is called the Troxler fading effect (Troxler, 1804). The vision of patients with complete oculomotor paralysis is thus probably very restricted to the center, if not close to blindness, except for very contrasted, rapidly changing and moving stimulations (e.g.: flashes used in visual evoked potentials). This is especially true in case of direct degeneration of oculomotor nuclei. In the case of frontal injury, it is possible that the preservation of reflexive oculomotor movement allows a slightly better perception. In a study were paralytic doses of curare were given to one participant, he reported the fading of the visual image (Stevens et al., 1976). Moreover, the ability to adjust the focus on the retina, namely accommodation, is also movement dependent: the ciliary muscle contraction. This muscle is innervated by the nerve III, which controls of most of our eye movements.

Motion is also required for accurate perception of most of the features processed by the vestibular, olfactory, gustative, tactile and proprioceptive systems. After only one week of hand and wrist immobilization, there is a disinhibition of the Ia afferent that conducts information along the muscle, which reflects a change in somatosensory inputs (Lundbye-Jensen & Nielsen, 2008). However, it is not mandatory that the motion is voluntarily generated. For example, passive texture perception is similar to active one (Heller, 1989; Lederman & Abbott, 1981; Simões-Franklin et al., 2011; Verrillo et al., 1999). The sense that is the most robust to the loss of voluntary motion is probably audition. However, the detection of the position of sounds in space can be improved with head motion (Coudert et al., 2022; Gaveau et al., 2022; Gessa et al., 2022; Pollack & Rose, 1967) and with eyes movements (Collins et al., 2010; Jones & Kabanoff, 1975).

### 2.3.2.2 Impaired motor execution impacts time and space perception

In auditory attention tasks with rhythmic patterns, participants that are allowed to beat time may present better attentional performance onto stimulations that are synchronized with their tapping, compared to participants that are asked to stay immobile (F. C. Manning et al., 2017; F. Manning & Schutz, 2013) (albeit this effect is not always found (London et al., 2019)). More generally, perception of time is influenced when performing movements (Yokosaka et al., 2015; Yon et al., 2017).

For example, slowing the movement using a viscous environment alters time perception (De Kock, Zhou, et al., 2021). Hence, persons in CLIS, deprived of macro and micro-movements, could present a decrease in temporal precision when trying to attend a stream of stimuli.

As put forward by Buzsáki, human organisms do not contain (as far as we know), direct sensors for time nor space (Buzsáki, 2019). He argues that, at least for short time durations, these parameters are inferred through the perception of the velocity of the organism motion, combined with the information of the position of the head and of the body. The hubs subserving the integration of these data are probably the parietal cortex and the hippocampal system, implicated both in time and space perception, which do not appear to be impacted directly from paralysis. However, the sensors that provide these data are located partly in the inner ear and in the proprioception system, and are much less informative when the subject is immobile. Otolith are one of these sensors, and recent findings uncovered their possible role in spatial cognition (P. F. Smith, 2019). Hence, this whole "spacetime" system could be altered by global paralysis. The term spacetime is used by Buzsáki both as closer to the fundamental state of the art of physics that do not separate time and space, but also to emphasize that it is possible that there are not two distinct systems for space and time in the brain, but one global feature from which the space and time can be derived from. This is parallel to the A Theory Of Magnitude (ATOM) proposed by Walsh (V. Walsh, 2003) that postulate a similar cortical processing for space and time. Research on space and time perception with persons with global paralysis is still in its infancy, and concern mainly the developmental aspects (Belmonti et al., 2016). Future research on precise time and space perception in complete paralysis could help to understand BCI performance, as most of BCI based on evoked potentials are based on spatial or temporal features.

### 2.3.2.3 Accurate body representation require motion?

Body representations have been typically decomposed into body schema and body image. In case of pure de-afferentation, the body schema seems to be mostly preserved, but it is no more based on proprioception but on body image, thanks to the visual feedback (de Vignemont, 2010). This is demonstrated by the fact that deafferented patients can move correctly as long as they can visually guide their limbs. However, a patient in CLIS has a double kind of loss: no motor action, no visual feedback of their body (or a very limited one). Hence their proprioception is poorly informative, and cannot be compensated by vision. Body schema and image of person suffering from LIS remain to be investigated, tough are likely to be impaired. There are some data in patients with spinal cord injuries showing a decrease in body ownership when the injury is higher (Lenggenhager et al., 2012).

There is evidence that some sensorial abilities such as peripheral vision are needed to achieve a precise spatial representation. Indeed, in patients with only central vision (without any impairment in motor control), there are impairments in spatial cognition, either allo- or ego-centered (Fortenbaugh et al., 2007, 2008; K. Turano, 1990; K. A. Turano et al., 2005; Yamamoto & Philbeck, 2013). Peripheral vision of patients with complete LIS is very likely to be impaired, because of oculomotor paralysis, which forces them to focus their attention on a space that they would misrepresent.

### 2.3.2.4 Impaired motor execution alters motor preparation and action selection

Studies with induced limb immobilization on healthy subjects (Newbold et al., 2020) show significant deleterious effects on both motor execution and motor imagery performances. These changes happen within days and even hours of immobilization. Disused motor regions became less connected to the opposite hemisphere and more strongly connected to each other, with strong spontaneous pulses coursing across these regions (Newbold et al., 2020, 2021). The functional role of these pulses is not known in these constrained immobilization situations, but they resemble the ones observed in developmental phases of the brain in preterm babies. Hence these pulses are supposed to play a role in the acquisition of coordination in development of the motor system and in "self-organization" of these neuronal populations (Khazipov et al., 2004; Tolonen et al., 2007).

There was no change in connectivity with the Default Mode Network (DMN), neither with the frontal nor salience network (Brauns et al., 2014; Debarnot et al., 2018; Fortuna et al., 2013; Lundbye-Jensen & Nielsen, 2008; Manaia et al., 2013; Schicatano et al., 2002). This could indicate that these disuse motor regions remain activated for some cognitive tasks.

### 2.3.2.5 Does impaired motor execution impact consciousness?

Immobilization leads to a lack of afferent information. This state can be approached in healthy subjects with Floatation-Reduced Environmental Stimulation. In this setting, subjects float in a controlled environment which reduces sensory stimulations. They report changes in their awareness that could be described as an intermediate state between sleep and wakefulness. Some subjects also report out of the body experiences. Some authors had participants undergoing an fMRI scan just after these experiences, and uncovered a reduction of connectivity within and between posterior hubs of the default mode network, and between the DMN and the primary and secondary somatomotor cortices (Al Zoubi et al., 2021). In pathological conditions without injury to the central nervous system, these kinds of experiences can be reported at the acute phase by one third of patients with a severe Guillain-barre syndrome (Cochen et al., 2005; Forsberg et al., 2008; Friedman et al., 2003; Liik et al., 2012). This condition induces a polyneuropathy leading to a brutal combination of de-afferentation and de-efferentation. The presence of such central neurological symptoms in a peripheral disease is not well explained. Some authors suggest that it is close to a sleep and dream associated disorder (Cochen et al., 2005). Hence, if the restriction of efferences and afferences does not lead to a loss of consciousness, it seems to at least alter the state of arousal in some subjects.

### 2.3.2.6 Conclusion

Impaired motor execution without direct injury of the central nervous system, even as short as a few hours, already impacts perception, motor preparation, action selection and arousal. The presence of such impairments and foremost their combination is likely to make the use of BCI more difficult. This may particularly apply to patients with complete paralysis, whose full immobilization is caused by the loss of integrity of the motor system. As we will see below, this is likely to induce other cognitive impairments.

## 2.4 Of the relationship between attention and motor control

### 1. Electrophysiological correlates of attention are strongly linked to action-related networks

At the electrophysiological level, markers of (covert) attention, like the P300, are less prevalent in patients with global paralysis (Lugo et al., 2016; Séguin et al., 2023; Wolpaw et al., 2018).

It is interesting to compare these results to recent literature on P300, which was simultaneously associated with attentional, exogenous and endogenous processes and action selection processes (Polich, 2007; Verleger et al., 2015). P300 is not *per se* a consciousness correlate, but can be considered as associated with an active task, reflecting the **link between stimulus and response** (Dembski et al., 2021; Verleger, 2020). For example, in a paradigm with stimulations around the discrimination threshold, the P300 was present only in the active task (Sergent et al., 2021). In the passive condition, the authors uncovered what they call "a global playground": associated with conscious perception only, and mainly confined to temporal lobe. Interestingly, the difference of activity between the "global playground" and the "global workspace" was mainly localized in task related network, as the motor cortex (Sergent et al., 2021).

Hence, the instability of P300 in global paralysis could reflect their difficulty to sustain an active task on stimuli. Consciousness would be preserved, but **the attempts of the task related networks to access the "global playground" and manipulate voluntarily the perceptual concepts** could be less efficient than in healthy subjects.

BCI paradigms can also rely on the attentional modulation of standard stimulations (i.e. not only the deviant ones): the endogenous amplification of one stimulation stream at the expense of another. These tasks require controlling both endogenous and exogenous attention, namely to inhibit distractors. In a steady-state evoked potential paradigm, Lesenfants et al. (Lesenfants et al., 2018) failed to detect a difference between attended and non-attended visual stimuli in patients with LIS. However, they could detect the effort of endogenous attention, i.e., the difference when subjects were doing the tasks versus passive reception of all stimulations. We analyzed the effect of attention on to-be-attended versus to-be-ignored standard sounds (Séguin et al., 2016) and found that only three out of seven patients with severe motor paralysis showed a modulation pattern (whereas 17 out of 18 healthy subjects did).

Could there be an impact of motor impairment on the attentional enhancement of stimulation? We gather here some findings about the role of the motor system during auditory and visual attentional tasks. Interestingly, a case report in an epileptic patient implanted with intra-cortical electrodes and presented with an oddball task uncovered an evoked activity for standard sounds in some frontal regions including three loci within the primary motor cortex (Kukleta et al., 2016). The patient was asked to mentally count targets, but no task was assigned for the standards, so either this late modulation of M1 was related to a change in motor preparation, or it plays a role in the processing of the sounds itself or of the rhythm of the sound. Moreover, as most BCI studies rely on concurrent streams of stimuli, it is informative to look at both the "enhancement" of target stimuli and the inhibition of the distracting ones. Distractors and targets appear to be encoded differently (van Moorselaar et al., 2020). Moreover, frequent and predictable distractors are more attenuated than rare ones, reflecting both habituation and statistical learning processes (for a review, see (Chelazzi et al., 2019) and (Vecera et al., 2014)). One electrophysiological marker of active inhibition is the distractor positivity (Pd) (Hickey et al., 2009), an ERP that happens contralaterally to the appearance of the distractor, around 200-300 ms after stimulus onset. In non-lateralized distractors, a global negative wave is observed (Daffner et al., 2012). A study with non-human primates uncovered that the homolog of the Pd wave appears 47 ms after a silencing of a population of neurons in a motor structure: the FEF. This effect can also be observed in the LIP (Ipata et al., 2006) and in the right middle frontal gyrus (Demeter et al., 2011; Marini et al., 2018). These findings are in line with the signal suppression hypothesis (Cosman et al., 2018). The population of FEF neurons that is tuned to the distractor strongly overlaps with the one that is more activated by the target. Hence, an impairment of FEF activity (as it is the case in ALS) probably has an impact on both target enhancement and target inhibition. Moreover, the sensory attenuation associated to self-motion that we described in previous part (2.4.2), is an example of how motor planning can induce a filtering of both interoceptive and exteroceptive information, and could help to potentiate voluntary selective attentional effort.

Overall, we hypothesize that 1) selective attentional effort is supported partly by both primary sensory and motor cortices 2) motor pathways play a role in the efficient maintenance of task-related stimuli in working memory. An impairment of the motor system could thus impair electrophysiological markers of attention.

### 2. *Is covert attention possible without (oculo-)motricity?*

As we saw previously, lack of oculomotor control dramatically affects visual perception. However, we could imagine that the subject remains able to focus their attention onto other type of sensory stimulations, covertly. The development of tactile or auditory gaze-independent BCI stems from this (non-verified) assumption.

The eyes are the first support of exploration of the environment for the newborn, and then the first support for the development of overt attentional abilities. Covert attention appears as soon as at 3 months of age in humans (Lunghi et al., 2020). However, covert attention is not so covert, at least at a micro-scale. Microsaccade are very fast movements happening at a rate of about 1 to 2 Hz. Different characteristics of microsaccades have been studied to understand their link with covert attention: direction, timing of onset and change of rate (including microsaccadic inhibition, which are a transient drop of microsaccade rate(Rolfs et al., 2008)). Microsaccades directions were shown to be correlated with the direction of covert attention (Hicheur et al., 2013; Lowet, Gomes, et al., 2018; Pastukhov & Braun, 2010; Yu et al., 2022; Yuval-Greenberg et al., 2014). An intra-cortical study in macaques found that the onset of attentional modulation was tightly coupled to microsaccades onset (Lowet, Gips, et al., 2018; Lowet, Gomes, et al., 2018). In oddball paradigms, salient events, even auditory ones, are associated with a temporary micro-saccades inhibition (Valsecchi et al., 2007; Valsecchi & Turatto, 2009; Widmann et al., 2014). The orientation or change of rate microsaccades are then most of the time a correlate of information processing, but their precise role remains to be explored (Yu et al., 2022).

It is remarkable that microsaccadic inhibition happens as early as 142 ms after stimulus onset. Some authors interpret this as a predictive information processing that makes some elaborated features available during early sensory processing at low level of sensory pathways (Widmann et al., 2014). This latter observation is also supported by invasive observation in human of very early voluntary attentional modulations in primary cortices, as early as 30 ms in the auditory cortex for example (Axelrod et al., 2022). This very early attentional processing has to rely on strong connections between motor and sensory pathways to be efficient, as each movement can induce changes in these sensory cortices. As global paralysis disrupts (almost all) these loops, there a possible effect on this early attentional processing.

Other movements of the eyes can be modulated by covert visuo-spatial attention: the angle of eye vergence increases when attending covertly a particular location in space that was previously indicated by either a visual or an auditory cue (Puig et al., 2013). This comes even earlier than another oculo-muscular correlate of attention: the pupillary dilatation (Strauch et al., 2022).

Covert attention is also solicited during mental imagery tasks. Even if attention is then oriented onto « internal » representations or body parts, instead of the outside world, healthy participants show eye movements that are specific to the tasks. For example, in mental imagery of spatial navigation, the eyes are moving as if the subject was exploring a real environment (Fourtassi et al., 2017). In motor imagery tasks, eye movements are close to those in motor execution, and dependent on the required movement amplitude (Heremans et al., 2008, 2011; Poiroux et al., 2015).

Are all these movements a consequence of the activation of similar networks between motricity and attention? Rizzolatti et al. postulated in 1987 that spatial attention is the consequence of activation of the premotor system (Rizzolatti et al., 1987). They proposed this framework based on the observation of the oculomotor system, especially during covert attention. They assumed an equivalence between the covert orienting of attention and the programming of explicit ocular movements. This theory is supported by the existence of pre-saccadic attention: when a saccade is prepared, attention is allocated to the future location of the gaze (Deubel & Schneider, 1996; Hoffman & Subramaniam, 1995). The fact that oculomotor mechanisms and visuospatial attention share similar cortical networks was confirmed by other studies (Beauchamp et al., 2001; Corbetta et al., 1998; Nobre et al., 2000).

Following theory of premotor attention, a prediction was that attention should be reduced to the oculomotor range, which was initially confirmed by several studies measuring reaction times (Craighero et al., 2001, 2004; D. T. Smith et al., 2004). However, recent studies from different groups reveal the possibility to pay attention beyond the limit of the oculomotor range (but inside the visual field), both for healthy subjects and patients with a selective oculomotor impairment, even when controlling for the head rotation range (Hanning et al., 2019; H.-H. Li, Pan, et al., 2021; Masson et al., 2020, 2021). Interestingly, the latter studies used a different behavioral marker of attentional performance, the discrimination threshold (instead of reaction time, which does not disentangle between detection and decision processes).

Hence, there is still currently an extensive debate on the link between attention and eye movements. Despite criticisms of some aspects of the premotor theory of attention, Smith and Schenk proposed in 2012 that a limited version of this theory was still valuable, namely the tight link between exogenous attention and motor preparation (D. T. Smith & Schenk, 2012). A recent review comparing the effects of covert attention and pre-saccadic attention showed that their mechanisms are different (H.-H. Li, Hanning, et al., 2021). The authors propose that these networks are close at a broad scale, encompassing the FEF and the precentral sulcus and the superior colliculus, but composed of different subpopulations of neurons.

Here, to answer our own question about the impact of global paralysis on attention, it is important to consider the different etiologies leading to these states. In ALS, when the disease reaches the FEF, it is not so selective, and can probably impair the different subpopulations of neurons, altering dramatically both covert and the residual pre-saccadic attention. For global paralysis due to brainstem injuries, the injuries principally alter the execution of the saccade. This is closer to the Moebius syndrome, which is a hypoplasia of the lower brainstem with impairment of oculomotor nerves inducing oculomotor paralysis. This syndrome was particularly studied in the framework of premotor theory of attention: when covert attention is assessed with reaction time, there is an impairment (Craighero et al., 2001), but when it is assessed with discrimination tasks, covert shift of attention are possible in the paralysed axis (Masson et al., 2020). Hence, as in the electrophysiological studies exposed previously, **it is possible that the impairment lies in the task-related networks, more than in the detection networks.**

Although inspired from the observation of the eyes, the premotor theory was also proposed more generally for any effectors. It was mainly about spatial attention, and did not make distinction between endogenous and exogenous attention. And indeed, attentional paradigms can also be applied to the limbs: it appears that when movements are preceded by an invalid motor cue, an increase in reaction times is observed (Brown et al., 2011). It is also possible to alter the programming and control of hand reaching movements using a visual attentional task (Aguilar Ros et al., 2021).

Besides, movement preparation without execution, is sufficient to generate sensory attenuation of the expected re-afferences (Voss et al., 2008; E. Walsh & Haggard, 2007). This sensory attenuation correlates with a reduced activity in the secondary somatosensory cortex and cerebellum (Blakemore et al., 1998), and an increased functional connectivity between these two areas (Blakemore et al., 1999; Kilteni et al., 2019). This sensory attenuation is not confined to somatosensory feedback, it is also observed when subjects trigger themselves some sounds by pressing a button (Martikainen et al., 2005). Furthermore, sensory attenuation is decreased when movements are involuntary (Bn et al., 2021). Altogether, this suggests that voluntary movement planning is helpful for strong sensory suppression. This sensory attenuation can be considered as part of a more global modulation of perception, allowing disentangling environmental from self-produced sensory inputs. Does this sensory attenuation rely on the same pathways as attention? Is it part of attention? We failed to find definitive answers to these questions in the literature, however there is a growing tendency in this field to confront this sensory attenuation to the rest of the field of perception and attention, especially in the light of predictive coding (Kaiser & Schütz-Bosbach, 2018; Kiepe et al., 2021; Kilteni & Ehrsson, 2022). In the most recent work, predictive attenuation of touch and tactile gating seem to be distinct perceptual phenomena (Kilteni & Ehrsson, 2022).Taken together, motor preparation is compatible with a top-down attentional process whose modulating or biasing effect can be mediated up to the lowest level of the motor hierarchy.

In practice, the divergence of motor preparation and attention orienting can induce delays in reaction times. This is in keeping with the fact that attention is mediated by a cross-modal network (Maravita et al., 2003) or a supra-modal network (C.-T. Wu et al., 2007). For instance, orienting attention toward a tactile target triggers an automatic displacement of spatial attention in the visual modality (Larson & Lee, 2013; C.-T. Wu et al., 2007). Likewise, in the frontal eye field (FEF), which orients the gaze and is involved in the planning of both covert and overt attention orienting, some neurons are directly activated by auditory cues (C.-T. Wu et al., 2007). Similarly, some areas in the primary motor cortex are modulated by attention, especially area 4p, which is highly connected with the primary sensory cortex (Binkofski et al., 2002; Gallego et al., 2022). Moreover, voluntary temporal attention toward a periodic stream of stimuli, either visual or auditory, strongly activates the inferior motor cortex in phase with the attended stream (Besle et al., 2011; Lakatos et al., 2008).

A rhythmic account of the relationship between motor activity and attention was proposed: sampling periods (with enhanced perception) would alternate with shifting periods (with overt or covert shifts of attention) (Fiebelkorn & Kastner, 2019; Anil Meera et al., 2022). Hence these networks are strongly linked, and working together, so the collapse of the motor network should have some impact on the attentional one, albeit it is difficult so far to know of what kind. Indeed, to our knowledge, the impact of the long-term impairment of overt shifts has not yet been explored.

### 3. *The risk of assuming that covert attention in severely paralyzed patients is spared*

Covert attention function in paralyzed patients appears to lack some of its fundamental markers, as compared to healthy participants: absence of micro-movements like microsaccades and poor quality of perception in almost all sensory modalities, especially in the spatial discrimination.

Moreover, processing of very salient events is different in patients. Auditory startle reflexes occur more frequently in case of stroke, but also in case of spinal cord injury (Abanoz et al., 2018; Jankelowitz & Colebatch, 2004; Kumru et al., 2008; S. Li et al., 2014). This hyper reactivity appears at some point in the post stroke recovery phase and is supposed to be correlated to the increase of activity in the reticulospinal tracts and reticular formation after injury (Choudhury et al., 2019; Fink & Cafferty, 2016).

ALS patients with residual voluntary oculomotor control present with abnormal reflexive oculomotor activity, in the form of saccadic intrusion that have a larger amplitude and duration (Becker et al., 2019).

Another difference between patients and healthy subjects in terms of covert attention stems from the quality of their multisensorial integration. Indeed, perception is usually obtained thanks to multiple senses (e.g. visual and proprioceptive), and the information are combined to minimize the uncertainty (van Beers et al., 2002). This is highlighted by different kinds of illusions. For example, in the rubber hand paradigm (Botvinick & Cohen, 1998), or in visuo-motor adaptation (Harris, 1963), an artificial discrepancy is generated between vision and proprioception. This uncover that there is an increased reliance (or weighting) on the spatially more precise modality, here on the visual one. These weights are flexible (Limanowski, 2021), and as an example, for in-depth perception, the proprioception can take the lead on vision because it is more precise in this axis (van Beers et al., 2002). Hence, after the onset of global paralysis, as the reliance of some senses decrease, there is probably a massive reshuffling of the weights of each sensory modality.

Chronic selective impairment of one sensory modality can probably be compensated for by other modalities, up to a certain extent. In case of severe motor paralysis, the somatosensory pathways become less available, as vestibular ones. It is very likely that more reliance is then attributed to the preserved pathways, such as vision when it is still preserved and audition. Due to this long-term shift, vision and audition probably support most of the attentional abilities at the locked-in stage, especially in the spatial modality. Hence, in case if there are additional oculomotor impairments, and patients become CLIS, they may have a dramatic impact on the spatial attentional span pertaining to these so far preserved sensory channels.

| Micro-movements that can be physiologically associated with attentional effort | Origin of Efferent | Possibles changes in case of global paralysis | References in physiological states | References in global paralysis | References in DOC |
|---|---|---|---|---|---|
| **Micro-saccades towards the localisation of attention** | Cranial nerve III, IV, VI | Absence in case of oculomotor paralysis, or pertubation due to saccadic intrusions or to nystagmus | (Widmann et al., 2014; Yu et al., 2022) | ? | ? |
| **Changes in pupillary size** | Cranial nerve III | Permanent mydriase in case of III injury | (Daniels et al., 2012; Hong et al., 2014; Strauch et al., 2022) | LIS: (Stoll et al., 2013) | (Stoll et al., 2013) |
| **Changes in vergences** | Cranial nerve III and IV | Absence if paralysis of the III | (Puig et al., 2013) | LIS: (Graber et al., 2016) | ? |
| **Changes in blink rate** | Cranial nerve III and VII | – Increase of blinks frequency in case of ALS<br>– Loss of eyelids control in case of advanced facial paralysis in ALS or direct injury on VII center in classical LIS | (Gusso et al., 2022) | ALS: (Byrne et al., 2013) | (Magliacano et al., 2021) |
| **Auriculomotor activity** |  | Change in basal muscle tone and lack of modulation during selective attention | (Strauss et al., 2020) | ? | ? |
| **Facial, Neck and trunk muscles contraction** |  | Change in basal muscle tone and lack of modulation during selective attention | (Jäncke et al., 1996) | ? | ? |
| **Respiration modulations** |  | No modulation if mechanical ventilation | (Park et al., 2020, 2022) | ? | (Arzi et al., 2020; Charland-Verville et al., 2014) |
| **Fingers muscles change in muscle tone** |  | Change in basal muscle tone and lack of modulation during selective attention | (Aravena et al., 2014; Boulenger et al., 2006) | ? | (Habbal et al., 2014) |

*Legend:*

*? : Lack of or very few data on this topic in the literature*

**Table 3:** *Involuntary micro-movements physiologically associated with covert attention and possible changes due to corticospinal injuries.*

# 3 Insights from general and computational theories

## 1. Role of motor control in theories of attention: from a pre-requisite to a constraint, toward a synonym?

Attention allows the selection of stimuli or thoughts, amongst all possible ones. The most influential and historical theories treat attention as a bottleneck which would be necessary because of the presumed limited ability of the brain to process all perceptual stimuli ("capacity-limitation theories") (Mole, 2017). This framework is historically anchored in a "passive perception" view, so the motor system is not at the core of these models. But knowing the effect of paralysis on perception, we can extrapolate some hypotheses. As we saw, in a total paralysis, there is a massive alteration of perception, in almost all sensory modalities. The inability to move and act upon the environment suppresses the need to accurately distinguish and select between objects. In this framework, a complete motor paralysis is not expected to affect attention per se, but the content of perception, reducing drastically the variety of accessible information about external objects and their salience, leading in the extreme case to the absence of perception (cf Troxler fading effect mentioned earlier). It is possible that because of this imbalance between poor external versus richer internal stimulations (e.g.: thoughts, dreams), there could be a tendency to increase attention to the richer "internal" world. In this case, attention would be dedicated to mind-wandering. This echoes the anti-correlation between "lateral" networks (turned toward external world) versus medial networks (turned toward internal thoughts). This view is also comparable to the one that is developed about the impact of sensory deprivation on thalamo-cortical pathways (Schmidt et al., 2020): in the lack of afferences, the outside world takes a back seat.

More recently, another group of theories of attention does not much emphasize its key role in dealing with limited resources. Amongst them, three streams are of particular relevance as they are related to each other and to question of interest: the "motor", the "selection for action" and the "precision optimization" theories of attention. Concerning the first group, we already mentioned the premotor theory of attention, according to which attention toward a particular location amounts to movement preparation in that direction. This would account for spatial attention impairment in case of paralysis, especially if the injury is on the premotor area or very connected with it. However, this theory has a limited explanatory power for non-spatial attention, feature-based attention for instance. We know that the detection of this type of attention can be found impaired in patients, for example when we ask them to focus on color (Lesenfants et al., 2018).

The "selection for action" theories of attention consider that the limitation, and then the need for attention, is due to a restricted possibility of actions, from bodily ones (Neisser, 1976) to every goal-directed action (Allport, 1987; O. Neumann, 1987). This is close to the view of Norman and Shallice, who proposed that "the primary role of attention is in the control of action" (Norman & Shallice, 1986), and introduced the concept of a "supervisory attentional system". Attention is then a necessity to allow a coherent course of action. In other words, what is limited is not our capacity of processing information, but our restricted range of actions. For Wu, attention plays a more important role for action than for perception, and attention is then strongly linked to agency and volition (Wu, 2011 ; Wu, 2019). The corollary of this point of view is that attention is crucial for managing the excess capacity: we can process too much stimuli and are at risk of distraction and lack of coherence in our behavior. For patients with global paralysis, the range of motor actions is abolished, but some mental actions seem to be preserved (e.g.: cognitive motor dissociation). Are these mental actions alone sufficient to maintain an efficient attention or is there some loss of attentional capacities? This hypothesis finds some echoes in others theories and in data obtained with patients in total paralysis. Patients in cognitive motor dissociation are known for being in a very fluctuant state (Chaudhary et al., 2022; Han et al., 2019), and they face an even more drastic drop of performance in self-initiated tasks that rely on their own volition (free communication mode) compared to more simple command following. Put together, this could be interpreted as a difficulty to maintain a coherent course of action and volition. This could also be due to an absence of spontaneous thinking activity, with only the ability to follow the input from environment.

Some computational theories of attention have tried to formalize mathematically the links between perception, attention and action, which provide more testable hypothesis. Precision-optimization theories arise from the bayesian inference framework (Vilares & Kording, 2011), which advocates that the brain is constantly building, testing and updating hypotheses or representations about the world. In the active inference view that formalizes these concepts in an enactive way (Brown et al., 2011), attention is considered as adjusting the weights of stimuli according to their expected reliance (precision) (Mirza et al., 2019; Parr & Friston, 2019). Recent developments within this framework made a distinction between attention and salience (Parr & Friston, 2019). Attention would be a gain adjustment toward current available and informative data. On the other hand, salience would be computed and assigned to data that have to be acquired in the future in order to best inform the hypothesis. Salience is thus tightly linked to the range of possible actions. Besides, the same framework also portrays (motor) action in terms of precision adjustment. Precisely, action is viewed a mean to minimize prediction error, by fulfilling sensory (exteroceptive and proprioceptive) predictions whose weights are also governed by precision parameters (Friston et al., 2010). Hence the dynamics of adjusting precision weights is sometimes referred to as (covert) actions and may share common anatomical and functional grounds with the ones of overt actions.

This brief theoretical overview is not intended to be exhaustive. Yet we covered a wide range of theories of attention, and in most of them, the motor system appears to play a prevalent role, either as a provider of new sensory information, as a constraint that act as a bottleneck, or as a similar process as attention (attention as a motor preparation or as precision optimization).

### 2. *Embodied, embedded, enactive, and extended (4E) cognition with a paralyzed body?*

Some authors with long experience in clinical BCI raised the question of a possible "extinction of goal directed thinking" that would accompany the occurrence of a CLIS (Kübler & Birbaumer, 2008). This concept was refined later by the same team in terms of "ideomotor silence", which would leads to a loss of voluntary responses and operant learning in long-term paralysis of human patients (Birbaumer et al., 2012; Birbaumer & Hummel, 2014). At the moment we lack empirical studies addressing directly this hypothesis which resonates with the more general theory of enaction (Varela, 1991, Kyselo & Di Paolo, 2015). The latter is also fully compatible with the empirical ascertainment that a strong reduction of the interaction with the environment (i.e., incomplete paralysis like LIS) has an impact on cognitive abilities.

Such an action-oriented view of cognition is currently experiencing a strong revival of interest (Buzsáki et al., 2014; Engel et al., 2013). It postulates that cognition should not be understood as providing models of the world, but as subsuming action and being grounded in sensorimotor coupling. These theories are part of the global concept of 4E cognition, that stands for embodied, embedded, enactive and extended cognition. As stated by Kyselo and Di Paolo, the common background of this vast and heterogeneous field might be synthetized with this sentence: "The *body is crucial for cognition*", but numerous discrepancies appear in the definition of what is "the body", what is "crucial" and what is "cognition" (Kyselo & Di Paolo, 2015). Considering the particular clinical cases that we address here; we question the role of the motor system: could the lack of voluntary control of skeletal muscles have a widespread impact on cognition?). A loss of integrity of the body corresponds here to injuries to the corticospinal tracts (first and second motoneurons), the nerves and their roots, and as a consequence, to a disconnection between brain and muscles. Albeit belonging to the central or peripheral nervous system, these parts are not considered as "cognitive" in the classical cognitivist view. One can argue that this situation does not strictly correspond to a "disembodiment" because the "classical body" is present, albeit the subject cannot move it voluntarily. Beyond the physical presence of the body, a crucial point here could be the ability of the patients to "feel" these paralyzed body parts as their own. This perception can probably vary between patients, especially with different etiologies that alters differentially somato-sensory pathways, including interoceptive ones. The completely "disembodied mind" would then be a patient with neither the ability to act nor to feel her/his body nor the environment, as described in (Bayne et al., 2020), but these cases are extremely rare.

A bunch of developmental literature reports that the motricity is highly important to acquire cognitive abilities (Libertus & Hauf, 2017). But as we saw, literature is sparser and more controversial about the impact of loss of motor control at an adult age. There seem to be a particular tipping point between the classical locked-in syndrome, where numerous abilities are preserved and cognitive-motor dissociation, where some responses to command are preserved, but with fluctuant reliability. Strikingly, it is difficult to find precise predictions in the 4E philosophical literature about the cognition in total paralysis. Kyselo and Di Paolo provide one of the rare works that address the theoretical challenges raised by the locked-in syndrome (Kyselo, 2020; Kyselo & Di Paolo, 2015). They compare three theories: sensorimotor approach, enactivism and extended cognition. They take as a starting point that the persons in LIS present with almost intact cognitive abilities, and then postulate that the theory that gives the best account of the LIS is the enactive framework (in accordance also with (Walter, 2010)). According to them, enactivism considers the body as an autonomous system and cognition as sense-making. Hence in the locked-in syndrome both are preserved, and embodiment is maintained. They predict that some abilities are more strongly embodied, such as spatial ones and social understandings, and then could be more impaired in LIS. On the contrary, and in opposition to our hypothesis and observations, they suppose that actions considered as "covert" like listening to speech should be more preserved. They do not study the particular case of complete locked-in state, which, to our knowledge, remains a blind spot in 4E literature, which is more dedicated to more extreme (and somehow caricatural) case like "the brain in a vat" problem. The brain of patients in complete locked-in state, if they ever exist, are not "in a vat", namely not completely disconnected from the rest of their body: there is a gut-brain axis, and most of the vegetative functions are preserved, as cardiac rhythms, blood pressure, stomach rhythms and respiration. Hence the need for dedicated theoretical studies of these complex cases remains.

# 4 Conclusion

Years of BCI attempts to help patients with severe motor impairment have revealed a paradox. These patients have long been regarded as cognitively intact and motivated enough to be the ideal candidates to benefit from neurotechnologies made for communication. However, performance of non-invasive BCI in these patients is lower than in healthy controls and no non-invasive BCI solution is routinely used by these patients who are left with no alternative. It urges the BCI community to consider the possible factors that beyond imperfect BCI systems and algorithms, may explain this striking observation.

Along all these years, BCI studies allowed to collect a significant and precious amount of electrophysiological data about these rare patients. This bunch of data questions the simplistic view of those patients as being purely motor impaired, with intact cognitive function. It turns out that whatever the etiology, cognitive impairments do come with motor ones, contradicting the view that cognition can function in disconnection from the body. On the other side, some radical positions in the field of embodied cognition posit the necessity of the body for cognition. Here, the existence of patients in cognitive-motor dissociation and in CLIS invites to nuance.

Our work emphasizes that the impact of paralysis on cognition is unlikely to be an "on-off" process. The big picture arising so far is more a combination of non-optimal functioning that converge toward an impairment of some functions, especially of attentional processes, and more generally of stable volitional effort. This adversary synergy could be overcome by some adaptive and compensation strategies, until a certain point.

Assessing these totally paralyzed patients require new technologies such as innovative brain-computer interfaces, but also more hypothesis driven paradigms to better explore their abilities. The development of computational theoretical accounts embracing the whole processes of action, perception and cognition is a promising and complementary pathway.

## Funding

PS was funded by two grants from the Fondation pour la Recherche Médicale (FRM, DEA20140629858 & FDM201906008524). PS and JM were also funded by the French government (ANR-17-CE40-0005, MindMadeClear). The COPHY team, IMPACT Team and CAP are supported by the Labex Cortex (ANR-11-LABX-0042).

Authors do not have any competing interest to declare.

# Bibliography

Abanoz, Y., Abanoz, Y., Gündüz, A., Uludağ, M., Örnek, N. İ., Uzun, N., Ünalan, H., & Kızıltan, M. (2018). Pattern of startle reflex to somatosensory stimuli changes after spinal cord injury. *The Journal of Spinal Cord Medicine*, *41*(1), 36-41. https://doi.org/10.1080/10790268.2016.1211580

Aguilar Ros, A., Mitchell, A. G., Ng, Y. W., & McIntosh, R. D. (2021). Attention attracts action in healthy participants : An insight into optic ataxia? *Cortex*, *137*, 149-159. https://doi.org/10.1016/j.cortex.2021.01.003

Al Zoubi, O., Misaki, M., Bodurka, J., Kuplicki, R., Wohlrab, C., Schoenhals, W. A., Refai, H. H., Khalsa, S. S., Stein, M. B., Paulus, M. P., & Feinstein, J. S. (2021). Taking the body off the mind : Decreased functional connectivity between somatomotor and default-mode networks following Floatation-REST. *Human Brain Mapping*, *42*(10), 3216-3227. https://doi.org/10.1002/hbm.25429

Allain, P., Joseph, P. A., Isambert, J. L., Le Gall, D., & Emile, J. (1998). Cognitive functions in chronic locked-in syndrome : A report of two cases. *Cortex; a Journal Devoted to the Study of the Nervous System and Behavior*, *34*(4), Article 4.

Allen, K. M., Lawlor, J., Salles, A., & Moss, C. F. (2021). Orienting our view of the superior colliculus : Specializations and general functions. *Current Opinion in Neurobiology*, *71*, 119-126. https://doi.org/10.1016/j.conb.2021.10.005

Allison, B. Z., Kübler, A., & Jin, J. (2020). 30+ years of P300 brain–computer interfaces. *Psychophysiology*, *57*(7), e13569. https://doi.org/10.1111/psyp.13569

Allport, A. (1987). Selection for Action : Some Behavioral and Neurophysiological Considerations of Attention and Action. In *Perspectives on Perception and Action*. Routledge.

Andres, M., Michaux, N., & Pesenti, M. (2012). Common substrate for mental arithmetic and finger representation in the parietal cortex. *NeuroImage*, *62*(3), 1520-1528. https://doi.org/10.1016/j.neuroimage.2012.05.047

Anil Meera, A., Novicky, F., Parr, T., Friston, K., Lanillos, P., & Sajid, N. (2022). Reclaiming saliency : Rhythmic precision-modulated action and perception. *Frontiers in Neurorobotics*, *16*, 896229. https://doi.org/10.3389/fnbot.2022.896229

Aravena, P., Courson, M., Frak, V., Cheylus, A., Paulignan, Y., Deprez, V., & Nazir, T. A. (2014). Action relevance in linguistic context drives word-induced motor activity. *Frontiers in Human Neuroscience*, *8*, 163. https://doi.org/10.3389/fnhum.2014.00163

Armstrong, K. M., Chang, M. H., & Moore, T. (2009). Selection and Maintenance of Spatial Information by Frontal Eye Field Neurons. *Journal of Neuroscience*, *29*(50), Article 50. https://doi.org/10.1523/JNEUROSCI.4465-09.2009

Arzi, A., Rozenkrantz, L., Gorodisky, L., Rozenkrantz, D., Holtzman, Y., Ravia, A., Bekinschtein, T. A., Galperin, T., Krimchansky, B.-Z., Cohen, G., Oksamitni, A., Aidinoff, E., Sacher, Y., & Sobel, N. (2020). Olfactory sniffing signals consciousness in unresponsive patients with brain injuries. *Nature*, *581*(7809), 428-433. https://doi.org/10.1038/s41586-020-2245-5

Aubinet, C., Chatelle, C., Gosseries, O., Carrière, M., Laureys, S., & Majerus, S. (2022). Residual implicit and explicit language abilities in patients with disorders of consciousness : A systematic review. *Neuroscience & Biobehavioral Reviews*, *132*, 391-409. https://doi.org/10.1016/j.neubiorev.2021.12.001

Avenanti, A., Annela, L., & Serino, A. (2012). Suppression of premotor cortex disrupts motor coding of peripersonal space. *NeuroImage*, *63*(1), 281-288. https://doi.org/10.1016/j.neuroimage.2012.06.063

Axelrod, V., Rozier, C., Lehongre, K., Adam, C., Lambrecq, V., Navarro, V., & Naccache, L. (2022). Neural modulations in the auditory cortex during internal and external attention tasks : A single-patient intracranial recording study. *Cortex*, *157*, 211-230. https://doi.org/10.1016/j.cortex.2022.09.011

Azzalini, D., Rebollo, I., & Tallon-Baudry, C. (2019). Visceral Signals Shape Brain Dynamics and Cognition. *Trends in Cognitive Sciences*, *23*(6), 488-509. https://doi.org/10.1016/j.tics.2019.03.007

Babiloni, C., Pistoia, F., Sarà, M., Vecchio, F., Buffo, P., Conson, M., Onorati, P., Albertini, G., & Rossini, P. M. (2010). Resting state eyes-closed cortical rhythms in patients with locked-in-syndrome : An EEG study. *Clinical Neurophysiology: Official Journal of the International Federation of Clinical Neurophysiology*, *121*(11), Article 11. https://doi.org/10.1016/j.clinph.2010.04.027

Bai, O., Lin, P., Vorbach, S., Floeter, M. K., Hattori, N., & Hallett, M. (2008). A high performance sensorimotor beta rhythm-based brain-computer interface associated with human natural motor behavior. *Journal of Neural Engineering*, *5*(1), Article 1. https://doi.org/10.1088/1741-2560/5/1/003

Barsalou, L. W. (2008). Grounded Cognition. *Annual Review of Psychology*, *59*(1), 617-645. https://doi.org/10.1146/annurev.psych.59.103006.093639

Bartolo, A., Carlier, M., Hassaini, S., Martin, Y., & Coello, Y. (2014). The perception of peripersonal space in right and left brain damage hemiplegic patients. *Frontiers in Human Neuroscience*, *8*. https://doi.org/10.3389/fnhum.2014.00003

Basaia, S., Agosta, F., Cividini, C., Trojsi, F., Riva, N., Spinelli, E. G., Moglia, C., Femiano, C., Castelnovo, V., Canu, E., Falzone, Y., Monsurrò, M. R., Falini, A., Chiò, A., Tedeschi, G., & Filippi, M. (2020). Structural and functional brain connectome in motor neuron diseases : A multicenter MRI study. *Neurology*, *95*(18), e2552-e2564. https://doi.org/10.1212/WNL.0000000000010731

Bassetti, C., Mathis, J., & Hess, C. W. (1994). Multimodal electrophysiological studies including motor evoked potentials in patients with locked-in syndrome : Report of six patients. *Journal of Neurology, Neurosurgery & Psychiatry*, *57*(11), 1403-1406. https://doi.org/10.1136/jnnp.57.11.1403

Basso, M. A., Bickford, M. E., & Cang, J. (2021). Unraveling circuits of visual perception and cognition through the superior colliculus. *Neuron*, *109*(6), 918-937. https://doi.org/10.1016/j.neuron.2021.01.013

Basso, M. A., & Wurtz, R. H. (1997). Modulation of neuronal activity by target uncertainty. *Nature*, *389*(6646), 66-69. https://doi.org/10.1038/37975

Bassolino, M., Finisguerra, A., Canzoneri, E., Serino, A., & Pozzo, T. (2015). Dissociating effect of upper limb non-use and overuse on space and body representations. *Neuropsychologia*, *70*, 385-392. https://doi.org/10.1016/j.neuropsychologia.2014.11.028

Bayne, T., Seth, A. K., & Massimini, M. (2020). Are There Islands of Awareness? *Trends in Neurosciences*, *43*(1), 6-16. https://doi.org/10.1016/j.tins.2019.11.003

Beauchamp, M. S., Petit, L., Ellmore, T. M., Ingeholm, J., & Haxby, J. V. (2001). A parametric fMRI study of overt and covert shifts of visuospatial attention. *NeuroImage*, *14*(2), Article 2. https://doi.org/10.1006/nimg.2001.0788

Becker, W., Gorges, M., Lulé, D., Pinkhardt, E., Ludolph, A. C., & Kassubek, J. (2019). Saccadic intrusions in amyotrophic lateral sclerosis (ALS). *Journal of Eye Movement Research*, *12*(6), Article 6. https://doi.org/10.16910/jemr.12.6.8

Bede, P., Chipika, R. H., Christidi, F., Hengeveld, J. C., Karavasilis, E., Argyropoulos, G. D., Lope, J., Shing, S. L. H., Velonakis, G., Dupuis, L., Doherty, M. A., Vajda, A., McLaughlin, R. L., & Hardiman, O. (2021). Genotype-associated cerebellar profiles in ALS : Focal cerebellar pathology and cerebro-cerebellar connectivity alterations. *Journal of Neurology, Neurosurgery & Psychiatry*, *92*(11), 1197-1205. https://doi.org/10.1136/jnnp-2021-326854

Bede, P., Elamin, M., Byrne, S., McLaughlin, R. L., Kenna, K., Vajda, A., Pender, N., Bradley, D. G., & Hardiman, O. (2013). Basal ganglia involvement in amyotrophic lateral sclerosis. *Neurology*, *81*(24), 2107-2115. https://doi.org/10.1212/01.wnl.0000437313.80913.2c

Beeldman, E., Raaphorst, J., Klein Twennaar, M., de Visser, M., Schmand, B. A., & de Haan, R. J. (2016). The cognitive profile of ALS : A systematic review and meta-analysis update. *Journal of Neurology, Neurosurgery, and Psychiatry*, *87*(6), Article 6. https://doi.org/10.1136/jnnp-2015-310734

Belmonti, V., Cioni, G., & Berthoz, A. (2016). Anticipatory control and spatial cognition in locomotion and navigation through typical development and in cerebral palsy. *Developmental Medicine and Child Neurology*, *58 Suppl 4*, 22-27. https://doi.org/10.1111/dmcn.13044

Bensch, M., Martens, S., Halder, S., Hill, J., Nijboer, F., Ramos, A., Birbaumer, N., Bogdan, M., Kotchoubey, B., Rosenstiel, W., Schölkopf, B., & Gharabaghi, A. (2014). Assessing attention and cognitive function in completely locked-in state with event-related brain potentials and epidural electrocorticography. *Journal of Neural Engineering*, *11*(2), Article 2. https://doi.org/10.1088/1741-2560/11/2/026006

Bernardi, L., Porta, C., & Sleight, P. (2006). Cardiovascular, cerebrovascular, and respiratory changes induced by different types of music in musicians and non-musicians : The importance of silence. *Heart (British Cardiac Society)*, *92*(4), 445-452. https://doi.org/10.1136/hrt.2005.064600

Besle, J., Schevon, C. A., Mehta, A. D., Lakatos, P., Goodman, R. R., McKhann, G. M., Emerson, R. G., & Schroeder, C. E. (2011). Tuning of the Human Neocortex to the Temporal Dynamics of Attended Events. *The Journal of Neuroscience*, *31*(9), 3176-3185. https://doi.org/10.1523/JNEUROSCI.4518-10.2011

Bibián, C., Irastorza-Landa, N., Schönauer, M., Birbaumer, N., López-Larraz, E., & Ramos-Murguialday, A. (2022). On the extraction of purely motor EEG neural correlates during an upper limb visuomotor task. *Cerebral Cortex*, *32*(19), 4243-4254. https://doi.org/10.1093/cercor/bhab479

Binkofski, F., Fink, G. R., Geyer, S., Buccino, G., Gruber, O., Shah, N. J., Taylor, J. G., Seitz, R. J., Zilles, K., & Freund, H.-J. (2002). Neural Activity in Human Primary Motor Cortex Areas 4a and 4p Is Modulated Differentially by Attention to Action. *Journal of Neurophysiology*, *88*(1), Article 1. https://doi.org/10.1152/jn.2002.88.1.514

Birbaumer, N. (2006). Breaking the silence : Brain-computer interfaces (BCI) for communication and motor control. *Psychophysiology*, *43*(6), Article 6. https://doi.org/10.1111/j.1469-8986.2006.00456.x

Birbaumer, N., & Hummel, F. C. (2014). Habit learning and brain–machine interfaces (BMI) : A tribute to Valentino Braitenberg's "Vehicles". *Biological Cybernetics*, *108*(5), Article 5. https://doi.org/10.1007/s00422-014-0595-5

Birbaumer, N., Piccione, F., Silvoni, S., & Wildgruber, M. (2012). Ideomotor silence : The case of complete paralysis and brain-computer interfaces (BCI). *Psychological Research*, *76*(2), Article 2. https://doi.org/10.1007/s00426-012-0412-5

Blakemore, S. J., Wolpert, D. M., & Frith, C. D. (1998). Central cancellation of self-produced tickle sensation. *Nature Neuroscience*, *1*(7), 635-640. https://doi.org/10.1038/2870

Blakemore, S. J., Wolpert, D. M., & Frith, C. D. (1999). The cerebellum contributes to somatosensory cortical activity during self-produced tactile stimulation. *NeuroImage*, *10*(4), 448-459. https://doi.org/10.1006/nimg.1999.0478

Blini, E., Desoche, C., Salemme, R., Kabil, A., Hadj-Bouziane, F., & Farnè, A. (2018). Mind the Depth : Visual Perception of Shapes Is Better in Peripersonal Space. *Psychological Science*, *29*(11), Article 11. https://doi.org/10.1177/0956797618795679

Bn, J., Mr, C., Rm, V., I, B., K, K.-B., Tj, W., & O, G. (2021). Movement Planning Determines Sensory Suppression : An Event-related Potential Study. *Journal of Cognitive Neuroscience*, *33*(12). https://doi.org/10.1162/jocn_a_01747

Bojorges-Valdez, E., Echeverría, J. C., & Yanez-Suarez, O. (2015). Evaluation of the continuous detection of mental calculation episodes as a BCI control input. *Computers in Biology and Medicine*, *64*, 155-162. https://doi.org/10.1016/j.compbiomed.2015.06.014

Bonnard, M., de Graaf, J., & Pailhous, J. (2004). Interactions between cognitive and sensorimotor functions in the motor cortex : Evidence from the preparatory motor sets anticipating a perturbation. *Reviews in the Neurosciences*, *15*(5), 371-382. https://doi.org/10.1515/revneuro.2004.15.5.371

Botvinick, M., & Cohen, J. (1998). Rubber hands 'feel' touch that eyes see. *Nature*, *391*(6669), Article 6669. https://doi.org/10.1038/35784

Boulenger, V., Roy, A. C., Paulignan, Y., Deprez, V., Jeannerod, M., & Nazir, T. A. (2006). Cross-talk between language processes and overt motor behavior in the first 200 msec of processing. *Journal of Cognitive Neuroscience*, *18*(10), 1607-1615. https://doi.org/10.1162/jocn.2006.18.10.1607

Brauns, I., Teixeira, S., Velasques, B., Bittencourt, J., Machado, S., Cagy, M., Gongora, M., Bastos, V. H., Machado, D., Sandoval-Carrillo, A., Salas-Pacheco, J., Piedade, R., Ribeiro, P., & Arias-Carrión, O. (2014). Changes in the theta band coherence during motor task after hand immobilization. *International Archives of Medicine*, *7*, 51. https://doi.org/10.1186/1755-7682-7-51

Brown, H., Friston, K., & Bestmann, S. (2011). Active inference, attention, and motor preparation. *Frontiers in Psychology*, *2*, 218. https://doi.org/10.3389/fpsyg.2011.00218

Brownell, B., Oppenheimer, D. R., & Hughes, J. T. (1970). The central nervous system in motor neurone disease. *Journal of Neurology, Neurosurgery, and Psychiatry*, *33*(3), Article 3.

Brozzoli, C., Ehrsson, H. H., & Farnè, A. (2014). Multisensory representation of the space near the hand : From perception to action and interindividual interactions. *The Neuroscientist: A Review Journal Bringing Neurobiology, Neurology and Psychiatry*, *20*(2), 122-135. https://doi.org/10.1177/1073858413511153

Brunner, P., Joshi, S., Briskin, S., Wolpaw, J. R., Bischof, H., & Schalk, G. (2010). Does the « P300 » speller depend on eye gaze? *Journal of Neural Engineering*, *7*(5), Article 5. https://doi.org/10.1088/1741-2560/7/5/056013

Bufacchi, R. J., & Iannetti, G. D. (2018). An Action Field Theory of Peripersonal Space. *Trends in Cognitive Sciences*, *22*(12), Article 12. https://doi.org/10.1016/j.tics.2018.09.004

Buzsáki, G. (2019). *The Brain from Inside Out*. Oxford University Press. https://doi.org/10.1093/oso/9780190905385.001.0001

Buzsáki, G., Peyrache, A., & Kubie, J. (2014). Emergence of Cognition from Action. *Cold Spring Harbor Symposia on Quantitative Biology*, *79*, 41-50. https://doi.org/10.1101/sqb.2014.79.024679

Byrne, S., Pradhan, F., Ni Dhubhghaill, S., Treacy, M., Cassidy, L., & Hardiman, O. (2013). Blink rate in ALS. *Amyotrophic Lateral Sclerosis & Frontotemporal Degeneration*, *14*(4), 291-293. https://doi.org/10.3109/21678421.2012.729217

Cappa, S. F., Pirovano, C., & Vignolo, L. A. (1985). Chronic « locked-in » syndrome : Psychological study of a case. *European Neurology*, *24*(2), Article 2. https://doi.org/10.1159/000115769

Cappa, S. F., & Vignolo, L. A. (1982). Locked-in syndrome for 12 years with preserved intelligence. *Annals of Neurology*, *11*(5), Article 5. https://doi.org/10.1002/ana.410110520

Charland-Verville, V., Lesenfants, D., Sela, L., Noirhomme, Q., Ziegler, E., Chatelle, C., Plotkin, A., Sobel, N., & Laureys, S. (2014). Detection of response to command using voluntary control of breathing in disorders of consciousness. *Frontiers in Human Neuroscience*, *8*, 1020. https://doi.org/10.3389/fnhum.2014.01020

Chaudhary, U., Vlachos, I., Zimmermann, J. B., Espinosa, A., Tonin, A., Jaramillo-Gonzalez, A., Khalili-Ardali, M., Topka, H., Lehmberg, J., Friehs, G. M., Woodtli, A., Donoghue, J. P., & Birbaumer, N. (2022). Spelling interface using intracortical signals in a completely locked-in patient enabled via auditory neurofeedback training. *Nature Communications*, *13*(1), 1236. https://doi.org/10.1038/s41467-022-28859-8

Chaudhary, U., Xia, B., Silvoni, S., Cohen, L. G., & Birbaumer, N. (2017). Brain-Computer Interface-Based Communication in the Completely Locked-In State. *PLoS Biology*, *15*(1), Article 1. https://doi.org/10.1371/journal.pbio.1002593

Chelazzi, L., Marini, F., Pascucci, D., & Turatto, M. (2019). Getting rid of visual distractors : The why, when, how, and where. *Current Opinion in Psychology*, *29*, 135-147. https://doi.org/10.1016/j.copsyc.2019.02.004

Choudhury, S., Shobhana, A., Singh, R., Sen, D., Anand, S. S., Shubham, S., Baker, M. R., Kumar, H., & Baker, S. N. (2019). The Relationship Between Enhanced Reticulospinal Outflow and Upper Limb Function in Chronic Stroke Patients. *Neurorehabilitation and Neural Repair*, *33*(5), 375-383. https://doi.org/10.1177/1545968319836233

Claassen, J., Doyle, K., Matory, A., Couch, C., Burger, K. M., Velazquez, A., Okonkwo, J. U., King, J.-R., Park, S., Agarwal, S., Roh, D., Megjhani, M., Eliseyev, A., Connolly, E. S., & Rohaut, B. (2019). Detection of Brain Activation in Unresponsive Patients with Acute Brain Injury. *New England Journal of Medicine*, *380*(26), Article 26. https://doi.org/10.1056/NEJMoa1812757

Cochen, V., Arnulf, I., Demeret, S., Neulat, M. L., Gourlet, V., Drouot, X., Moutereau, S., Derenne, J. P., Similowski, T., Willer, J. C., Pierrot-Deseiligny, C., & Bolgert, F. (2005). Vivid dreams, hallucinations, psychosis and REM sleep in Guillain–Barré syndrome. *Brain*, *128*(11), 2535-2545. https://doi.org/10.1093/brain/awh585

Collins, T., Heed, T., & Röder, B. (2010). Eye-movement-driven changes in the perception of auditory space. *Attention, Perception, & Psychophysics*, *72*(3), 736-746. https://doi.org/10.3758/APP.72.3.736

Conson, M., Pistoia, F., Sarà, M., Grossi, D., & Trojano, L. (2010). Recognition and mental manipulation of body parts dissociate in locked-in syndrome. *Brain and Cognition*, *73*(3), Article 3. https://doi.org/10.1016/j.bandc.2010.05.001

Conson, M., Sacco, S., Sarà, M., Pistoia, F., Grossi, D., & Trojano, L. (2008). Selective motor imagery defect in patients with locked-in syndrome. *Neuropsychologia*, *46*(11), Article 11. https://doi.org/10.1016/j.neuropsychologia.2008.04.015

Corbetta, M., Akbudak, E., Conturo, T. E., Snyder, A. Z., Ollinger, J. M., Drury, H. A., Linenweber, M. R., Petersen, S. E., Raichle, M. E., Van Essen, D. C., & Shulman, G. L. (1998). A common network of functional areas for attention and eye movements. *Neuron*, *21*(4), Article 4.

Cosman, J. D., Lowe, K. A., Woodman, G. F., & Schall, J. D. (2018). Prefrontal control of visual distraction. *Current biology : CB*, *28*(3), 414-420.e3. https://doi.org/10.1016/j.cub.2017.12.023

Coudert, A., Gaveau, V., Gatel, J., Verdelet, G., Salemme, R., Farne, A., Pavani, F., & Truy, E. (2022). Spatial Hearing Difficulties in Reaching Space in Bilateral Cochlear Implant Children Improve With Head Movements. *Ear and Hearing*, *43*(1), 192-205. https://doi.org/10.1097/AUD.0000000000001090

Craighero, L., Carta, A., & Fadiga, L. (2001). Peripheral oculomotor palsy affects orienting of visuospatial attention. *Neuroreport*, *12*(15), Article 15.

Craighero, L., Nascimben, M., & Fadiga, L. (2004). Eye position affects orienting of visuospatial attention. *Current Biology: CB*, *14*(4), Article 4. https://doi.org/10.1016/j.cub.2004.01.054

Criscuolo, A., Schwartze, M., & Kotz, S. A. (2022). Cognition through the lens of a body–brain dynamic system. *Trends in Neurosciences*, *0*(0). https://doi.org/10.1016/j.tins.2022.06.004

Cruse, D., Chennu, S., Chatelle, C., Bekinschtein, T. A., Fernández-Espejo, D., Pickard, J. D., Laureys, S., & Owen, A. M. (2011). Bedside detection of awareness in the vegetative state : A cohort study. *The Lancet*, *378*(9809), Article 9809. https://doi.org/10.1016/S0140-6736(11)61224-5

Daffner, K. R., Zhuravleva, T. Y., Sun, X., Tarbi, E. C., Haring, A. E., Rentz, D. M., & Holcomb, P. J. (2012). Does modulation of selective attention to features reflect enhancement or suppression of neural activity? *Biological Psychology*, *89*(2), 398-407. https://doi.org/10.1016/j.biopsycho.2011.12.002

Daniels, L. B., Nichols, D. F., Seifert, M. S., & Hock, H. S. (2012). Changes in pupil diameter entrained by cortically initiated changes in attention. *Visual Neuroscience*, *29*(2), 131-142. https://doi.org/10.1017/S0952523812000077

Davidson, M. J., Mithen, W., Hogendoorn, H., van Boxtel, J. J., & Tsuchiya, N. (2020). The SSVEP tracks attention, not consciousness, during perceptual filling-in. *eLife*, *9*, e60031. https://doi.org/10.7554/eLife.60031

De Kock, R., Gladhill, K. A., Ali, M. N., Joiner, W. M., & Wiener, M. (2021). How movements shape the perception of time. *Trends in Cognitive Sciences*, *25*(11), 950-963. https://doi.org/10.1016/j.tics.2021.08.002

De Kock, R., Zhou, W., Joiner, W. M., & Wiener, M. (2021). Slowing the body slows down time perception. *eLife*, *10*, e63607. https://doi.org/10.7554/eLife.63607

Debarnot, U., Huber, C., Guillot, A., & Schwartz, S. (2018). Sensorimotor representation and functional motor changes following short-term arm immobilization. *Behavioral Neuroscience*, *132*(6), 595-603. https://doi.org/10.1037/bne0000274

Decety, J., Jeannerod, M., Durozard, D., & Baverel, G. (1993). Central activation of autonomic effectors during mental simulation of motor actions in man. *The Journal of Physiology*, *461*(1), 549-563. https://doi.org/10.1113/jphysiol.1993.sp019528

Dembski, C., Koch, C., & Pitts, M. (2021). *Physiological Correlates of Sensory Consciousness : Evidence for a Perceptual Awareness Negativity*. PsyArXiv. https://doi.org/10.31234/osf.io/ma36g

Demeter, E., Hernandez-Garcia, L., Sarter, M., & Lustig, C. (2011). Challenges to attention : A continuous arterial spin labeling (ASL) study of the effects of distraction on sustained attention. *NeuroImage*, *54*(2), 1518-1529. https://doi.org/10.1016/j.neuroimage.2010.09.026

Deubel, H., & Schneider, W. X. (1996). Saccade target selection and object recognition : Evidence for a common attentional mechanism. *Vision Research*, *36*(12), 1827-1837. https://doi.org/10.1016/0042-6989(95)00294-4

de Vignemont, F. (2010). Body schema and body image—Pros and cons. *Neuropsychologia*, *48*(3), 669-680. https://doi.org/10.1016/j.neuropsychologia.2009.09.022

Di Paolo, E. (2019). Process and Individuation : The Development of Sensorimotor Agency. *Human Development*, 1-25. https://doi.org/10.1159/000503827

Dipoppa, M., Ranson, A., Krumin, M., Pachitariu, M., Carandini, M., & Harris, K. D. (2018). Vision and Locomotion Shape the Interactions between Neuron Types in Mouse Visual Cortex. *Neuron*, *98*(3), 602-615.e8. https://doi.org/10.1016/j.neuron.2018.03.037

Editors, T. P. B. (2019). Retraction : Response to: "Questioning the evidence for BCI-based communication in the complete locked-in state". *PLOS Biology*, *17*(12), Article 12. https://doi.org/10.1371/journal.pbio.3000608

Egbebike, J., Shen, Q., Doyle, K., Der-Nigoghossian, C. A., Panicker, L., Gonzales, I. J., Grobois, L., Carmona, J. C., Vrosgou, A., Kaur, A., Boehme, A., Velazquez, A., Rohaut, B., Roh, D., Agarwal, S., Park, S., Connolly, E. S., & Claassen, J. (2022). Cognitive-motor dissociation and time to functional recovery in patients with acute brain injury in the USA : A prospective observational cohort study. *The Lancet Neurology*, *21*(8), 704-713. https://doi.org/10.1016/S1474-4422(22)00212-5

Engel, A. K., Maye, A., Kurthen, M., & König, P. (2013). Where's the action? The pragmatic turn in cognitive science. *Trends in Cognitive Sciences*, *17*(5), Article 5. https://doi.org/10.1016/j.tics.2013.03.006

Erisken, S., Vaiceliunaite, A., Jurjut, O., Fiorini, M., Katzner, S., & Busse, L. (2014). Effects of locomotion extend throughout the mouse early visual system. *Current Biology: CB*, *24*(24), 2899-2907. https://doi.org/10.1016/j.cub.2014.10.045

Faskowitz, J., Esfahlani, F. Z., Jo, Y., Sporns, O., & Betzel, R. F. (2020). Edge-centric functional network representations of human cerebral cortex reveal overlapping system-level architecture. *Nature Neuroscience*, *23*(12), Article 12. https://doi.org/10.1038/s41593-020-00719-y

Fernández-Espejo, D., & Owen, A. M. (2013). Detecting awareness after severe brain injury. *Nature Reviews Neuroscience*, *14*(11), Article 11. https://doi.org/10.1038/nrn3608

Ferré, F., Heine, L., Naboulsi, E., Gobert, F., Beaudoin-Gobert, M., Dailler, F., Buffières, W., Corneyllie, A., Sarton, B., Riu, B., Luauté, J., Silva, S., & Perrin, F. (2023). Self-processing in coma, unresponsive wakefulness syndrome and minimally conscious state. *Frontiers in Human Neuroscience*, *17*. https://www.frontiersin.org/articles/10.3389/fnhum.2023.1145253

Fiebelkorn, I. C., & Kastner, S. (2019). A Rhythmic Theory of Attention. *Trends in Cognitive Sciences*, *23*(2), 87-101. https://doi.org/10.1016/j.tics.2018.11.009

Fink, K. L., & Cafferty, W. B. J. (2016). Reorganization of Intact Descending Motor Circuits to Replace Lost Connections After Injury. *Neurotherapeutics*, *13*(2), 370-381. https://doi.org/10.1007/s13311-016-0422-x

Forsberg, A., Ahlström, G., & Holmqvist, L. W. (2008). Falling ill with Guillain-Barré syndrome : Patients' experiences during the initial phase. *Scandinavian Journal of Caring Sciences*, *22*(2), 220-226. https://doi.org/10.1111/j.1471-6712.2007.00517.x

Fortenbaugh, F. C., Hicks, J. C., Hao, L., & Turano, K. A. (2007). Losing sight of the bigger picture : Peripheral field loss compresses representations of space. *Vision Research*, *47*(19), 2506-2520. https://doi.org/10.1016/j.visres.2007.06.012

Fortenbaugh, F. C., Hicks, J. C., & Turano, K. A. (2008). The effect of peripheral visual field loss on representations of space : Evidence for distortion and adaptation. *Investigative Ophthalmology & Visual Science*, *49*(6), 2765-2772. https://doi.org/10.1167/iovs.07-1021

Fortuna, M., Teixeira, S., Machado, S., Velasques, B., Bittencourt, J., Peressutti, C., Budde, H., Cagy, M., Nardi, A. E., Piedade, R., Ribeiro, P., & Arias-Carrión, O. (2013). Cortical reorganization after hand immobilization : The beta qEEG spectral coherence evidences. *PloS One*, *8*(11), e79912. https://doi.org/10.1371/journal.pone.0079912

Fourtassi, M., Rode, G., & Pisella, L. (2017). Using eye movements to explore mental representations of space. *Annals of Physical and Rehabilitation Medicine*, *60*(3), 160-163. https://doi.org/10.1016/j.rehab.2016.03.001

Freedman, D. J., & Ibos, G. (2018). An Integrative Framework for Sensory, Motor, and Cognitive Functions of the Posterior Parietal Cortex. *Neuron*, *97*(6), Article 6. https://doi.org/10.1016/j.neuron.2018.01.044

Freudenburg, Z. V., Branco, M. P., Leinders, S., Vijgh, B. H. van der, Pels, E. G. M., Denison, T., Berg, L. H. van den, Miller, K. J., Aarnoutse, E. J., Ramsey, N. F., & Vansteensel, M. J. (2019). Sensorimotor ECoG Signal Features for BCI Control : A Comparison Between People With Locked-In Syndrome and Able-Bodied Controls. *Frontiers in Neuroscience*, *13*. https://doi.org/10.3389/fnins.2019.01058

Friedman, Y., Lee, L., Wherrett, J. R., Ashby, P., & Carpenter, S. (2003). Simulation of Brain Death from Fulminant De-efferentation. *Canadian Journal of Neurological Sciences*, *30*(4), 397-404. https://doi.org/10.1017/S0317167100003152

Friston, K. J., Daunizeau, J., Kilner, J., & Kiebel, S. J. (2010). Action and behavior : A free-energy formulation. *Biological Cybernetics*, *102*(3), 227-260. https://doi.org/10.1007/s00422-010-0364-z

Gallego, J. A., Makin, T. R., & McDougle, S. D. (2022). Going beyond primary motor cortex to improve brain-computer interfaces. *Trends in Neurosciences*, *45*(3), 176-183. https://doi.org/10.1016/j.tins.2021.12.006

Gaveau, V., Coudert, A., Salemme, R., Koun, E., Desoche, C., Truy, E., Farnè, A., & Pavani, F. (2022). Benefits of active listening during 3D sound localization. *Experimental Brain Research*, *240*(11), 2817-2833. https://doi.org/10.1007/s00221-022-06456-x

Gayraud, F., Martinie, B., Bentot, E., Lepilliez, A., Tell, L., Cotton, F., & Rode, G. (2015). Written production in a case of locked-in syndrome with bilateral corticopontic degeneration. *Neuropsychological Rehabilitation*, *25*(5), Article 5. https://doi.org/10.1080/09602011.2014.975253

Gessa, E., Giovanelli, E., Spinella, D., Verdelet, G., Farnè, A., Frau, G. N., Pavani, F., & Valzolgher, C. (2022). Spontaneous head-movements improve sound localization in aging adults with hearing loss. *Frontiers in Human Neuroscience*, *16*, 1026056. https://doi.org/10.3389/fnhum.2022.1026056

Graber, M., Challe, G., Alexandre, M. F., Bodaghi, B., LeHoang, P., & Touitou, V. (2016). Evaluation of the visual function of patients with locked-in syndrome : Report of 13 cases. *Journal Francais D'ophtalmologie*, *39*(5), Article 5. https://doi.org/10.1016/j.jfo.2016.01.005

Guger, C., Spataro, R., Allison, B. Z., Heilinger, A., Ortner, R., Cho, W., & La Bella, V. (2017). Complete Locked-in and Locked-in Patients : Command Following Assessment and Communication with Vibro-Tactile P300 and Motor Imagery Brain-Computer Interface Tools. *Frontiers in Neuroscience*, *11*, 251. https://doi.org/10.3389/fnins.2017.00251

Gusso, M. M., Christison-Lagay, K. L., Zuckerman, D., Chandrasekaran, G., Kronemer, S. I., Ding, J. Z., Freedman, N. C., Nohama, P., & Blumenfeld, H. (2022). More than a feeling : Scalp EEG and eye signals in conscious tactile perception. *Consciousness and Cognition*, *105*, 103411. https://doi.org/10.1016/j.concog.2022.103411

Haas, F., Distenfeld, S., & Axen, K. (1986). Effects of perceived musical rhythm on respiratory pattern. *Journal of Applied Physiology*, *61*(3), 1185-1191. https://doi.org/10.1152/jappl.1986.61.3.1185

Habbal, D., Gosseries, O., Noirhomme, Q., Renaux, J., Lesenfants, D., Bekinschtein, T. A., Majerus, S., Laureys, S., & Schnakers, C. (2014). Volitional electromyographic responses in disorders of consciousness. *Brain Injury*, *28*(9), 1171-1179. https://doi.org/10.3109/02699052.2014.920519

Haggard, P. (2017). Sense of agency in the human brain. *Nature Reviews. Neuroscience*, *18*(4), 196-207. https://doi.org/10.1038/nrn.2017.14

Han, C.-H., Kim, Y.-W., Kim, D. Y., Kim, S. H., Nenadic, Z., & Im, C.-H. (2019). Electroencephalography-based endogenous brain-computer interface for online communication with a completely locked-in patient. *Journal of Neuroengineering and Rehabilitation*, *16*(1), Article 1. https://doi.org/10.1186/s12984-019-0493-0

Hanning, N. M., Szinte, M., & Deubel, H. (2019). Visual attention is not limited to the oculomotor range. *Proceedings of the National Academy of Sciences*, *116*(19), Article 19. https://doi.org/10.1073/pnas.1813465116

Harris, C. S. (1963). Adaptation to Displaced Vision : Visual, Motor, or Proprioceptive Change? *Science*, *140*(3568), 812-813. https://doi.org/10.1126/science.140.3568.812

Hayashi, K., Mochizuki, Y., Nakayama, Y., Shimizu, T., Kawata, A., Nagao, M., Mizutani, T., & Matsubara, S. (2013). [Communication disorder in amyotrophic lateral sclerosis after ventilation—A proposal of staging and a study of predictive factor]. *Rinshō shinkeigaku = Clinical neurology*, *53*(2), Article 2.

Heller, M. A. (1989). Texture perception in sighted and blind observers. *Perception & Psychophysics*, *45*(1), 49-54. https://doi.org/10.3758/bf03208032

Heremans, E., Helsen, W. F., & Feys, P. (2008). The eyes as a mirror of our thoughts : Quantification of motor imagery of goal-directed movements through eye movement registration. *Behavioural Brain Research*, *187*(2), 351-360. https://doi.org/10.1016/j.bbr.2007.09.028

Heremans, E., Smits-Engelsman, B., Caeyenberghs, K., Vercruysse, S., Nieuwboer, A., Feys, P., & Helsen, W. F. (2011). Keeping an eye on imagery : The role of eye movements during motor imagery training. *Neuroscience*, *195*, 37-44. https://doi.org/10.1016/j.neuroscience.2011.07.030

Herrero, J. L., Khuvis, S., Yeagle, E., Cerf, M., & Mehta, A. D. (2018). Breathing above the brain stem : Volitional control and attentional modulation in humans. *Journal of Neurophysiology*, *119*(1), 145-159. https://doi.org/10.1152/jn.00551.2017

Hicheur, H., Zozor, S., Campagne, A., & Chauvin, A. (2013). Microsaccades are modulated by both attentional demands of a visual discrimination task and background noise. *Journal of vision*. https://doi.org/10.1167/13.13.18

Hickey, C., Di Lollo, V., & McDonald, J. J. (2009). Electrophysiological Indices of Target and Distractor Processing in Visual Search. *Journal of Cognitive Neuroscience*, *21*(4), 760-775. https://doi.org/10.1162/jocn.2009.21039

Hoffman, J. E., & Subramaniam, B. (1995). The role of visual attention in saccadic eye movements. *Perception & Psychophysics*, *57*(6), 787-795. https://doi.org/10.3758/bf03206794

Holz, E. M., Botrel, L., Kaufmann, T., & Kübler, A. (2015). Long-term independent brain-computer interface home use improves quality of life of a patient in the locked-in state : A case study. *Archives of Physical Medicine and Rehabilitation*, *96*(3 Suppl), Article 3 Suppl. https://doi.org/10.1016/j.apmr.2014.03.035

Hong, L., Walz, J. M., & Sajda, P. (2014). Your Eyes Give You Away : Prestimulus Changes in Pupil Diameter Correlate with Poststimulus Task-Related EEG Dynamics. *PLoS ONE*, *9*(3), e91321. https://doi.org/10.1371/journal.pone.0091321

Hou, S., & Rabchevsky, A. G. (2014). Autonomic consequences of spinal cord injury. *Comprehensive Physiology*, *4*(4), 1419-1453. https://doi.org/10.1002/cphy.c130045

Hudson, A. J. (1981). Amyotrophic lateral sclerosis and its association with dementia, parkinsonism and other neurological disorders : A review. *Brain: A Journal of Neurology*, *104*(2), Article 2.

Iglesias, C., Sangari, S., El Mendili, M.-M., Benali, H., Marchand-Pauvert, V., & Pradat, P.-F. (2015). Electrophysiological and spinal imaging evidences for sensory dysfunction in amyotrophic lateral sclerosis. *BMJ Open*, *5*(2), e007659. https://doi.org/10.1136/bmjopen-2015-007659

Ignashchenkova, A., Dicke, P. W., Haarmeier, T., & Thier, P. (2004). Neuron-specific contribution of the superior colliculus to overt and covert shifts of attention. *Nature Neuroscience*, *7*(1), 56-64. https://doi.org/10.1038/nn1169

Ipata, A. E., Gee, A. L., Gottlieb, J., Bisley, J. W., & Goldberg, M. E. (2006). LIP responses to a popout stimulus are reduced if it is overtly ignored. *Nature Neuroscience*, *9*(8), Article 8. https://doi.org/10.1038/nn1734

Isaacs, J. D., Dean, A. F., Shaw, C. E., Al-Chalabi, A., Mills, K. R., & Leigh, P. N. (2007). Amyotrophic lateral sclerosis with sensory neuropathy : Part of a multisystem disorder? *Journal of Neurology, Neurosurgery, and Psychiatry*, *78*(7), Article 7. https://doi.org/10.1136/jnnp.2006.098798

Janacsek, K., Evans, T. M., Kiss, M., Shah, L., Blumenfeld, H., & Ullman, M. T. (2022). Subcortical Cognition : The Fruit Below the Rind. *Annual Review of Neuroscience*, *45*, 361-386. https://doi.org/10.1146/annurev-neuro-110920-013544

Jäncke, L., Vogt, J., Musial, F., Lutz, K., & Kalveram, K. T. (1996). Facial EMG responses to auditory stimuli. *International Journal of Psychophysiology: Official Journal of the International Organization of Psychophysiology*, *22*(1-2), 85-96. https://doi.org/10.1016/0167-8760(96)00013-x

Jankelowitz, S. K., & Colebatch, J. G. (2004). The acoustic startle reflex in ischemic stroke. *Neurology*, *62*(1), 114-116. https://doi.org/10.1212/01.wnl.0000101711.48946.35

Jeon, H., & Shin, D. A. (2015). Experimental Set Up of P300 Based Brain Computer Interface Using a Bioamplifier and BCI2000 System for Patients with Spinal Cord Injury. *Korean Journal of Spine*, *12*(3), Article 3. https://doi.org/10.14245/kjs.2015.12.3.119

Jones, B., & Kabanoff, B. (1975). Eye movements in auditory space perception. *Perception & Psychophysics*, *17*(3), 241-245. https://doi.org/10.3758/BF03203206

Kaiser, J., & Schütz-Bosbach, S. (2018). Sensory attenuation of self-produced signals does not rely on self-specific motor predictions. *The European Journal of Neuroscience*, *47*(11), 1303-1310. https://doi.org/10.1111/ejn.13931

Kaufman, M. T., Churchland, M. M., Ryu, S. I., & Shenoy, K. V. (2014). Cortical activity in the null space : Permitting preparation without movement. *Nature Neuroscience*, *17*(3), Article 3. https://doi.org/10.1038/nn.3643

Kaufmann, T., Holz, E. M., & Kübler, A. (2013). Comparison of tactile, auditory, and visual modality for brain-computer interface use : A case study with a patient in the locked-in state. *Frontiers in Neuroscience*, *7*, 129. https://doi.org/10.3389/fnins.2013.00129

Kellmeyer, P., Grosse-Wentrup, M., Schulze-Bonhage, A., Ziemann, U., & Ball, T. (2018). Electrophysiological correlates of neurodegeneration in motor and non-motor brain regions in amyotrophic lateral sclerosis—Implications for brain–computer interfacing. *Journal of Neural Engineering*, *15*(4), 041003. https://doi.org/10.1088/1741-2552/aabfa5

Khazipov, R., Sirota, A., Leinekugel, X., Holmes, G. L., Ben-Ari, Y., & Buzsáki, G. (2004). Early motor activity drives spindle bursts in the developing somatosensory cortex. *Nature*, *432*(7018), 758-761. https://doi.org/10.1038/nature03132

Kiepe, F., Kraus, N., & Hesselmann, G. (2021). Sensory Attenuation in the Auditory Modality as a Window Into Predictive Processing. *Frontiers in Human Neuroscience*, *15*. https://www.frontiersin.org/articles/10.3389/fnhum.2021.704668

Kilteni, K., & Ehrsson, H. H. (2022). Predictive attenuation of touch and tactile gating are distinct perceptual phenomena. *iScience*, *25*(4), 104077. https://doi.org/10.1016/j.isci.2022.104077

Kilteni, K., Houborg, C., & Ehrsson, H. H. (2019). Rapid learning and unlearning of predicted sensory delays in self-generated touch. *eLife*, *8*, e42888. https://doi.org/10.7554/eLife.42888

Klein, C., & Ettinger, U. (Éds.). (2019). *Eye Movement Research : An Introduction to its Scientific Foundations and Applications*. Springer International Publishing. https://doi.org/10.1007/978-3-030-20085-5

Kluger, D. S., Balestrieri, E., Busch, N. A., & Gross, J. (2021). Respiration aligns perception with neural excitability. *eLife*, *10*, e70907. https://doi.org/10.7554/eLife.70907

Kodaka, Y., Mikami, A., & Kubota, K. (1997). Neuronal activity in the frontal eye field of the monkey is modulated while attention is focused on to a stimulus in the peripheral visual field, irrespective of eye movement. *Neuroscience Research*, *28*(4), Article 4. https://doi.org/10.1016/S0168-0102(97)00055-2

Kondziella, D., Friberg, C. K., Frokjaer, V. G., Fabricius, M., & Møller, K. (2016). Preserved consciousness in vegetative and minimal conscious states : Systematic review and meta-analysis. *Journal of Neurology, Neurosurgery & Psychiatry*, *87*(5), 485-492. https://doi.org/10.1136/jnnp-2015-310958

Krauzlis, R. J., Bogadhi, A. R., Herman, J. P., & Bollimunta, A. (2018). Selective attention without a neocortex. *Cortex*, *102*, 161-175. https://doi.org/10.1016/j.cortex.2017.08.026

Kübler, A., & Birbaumer, N. (2008). Brain–computer interfaces and communication in paralysis : Extinction of goal directed thinking in completely paralysed patients? *Clinical Neurophysiology*, *119*(11), Article 11. https://doi.org/10.1016/j.clinph.2008.06.019

Kübler, A., Holz, E. M., Riccio, A., Zickler, C., Kaufmann, T., Kleih, S. C., Staiger-Sälzer, P., Desideri, L., Hoogerwerf, E.-J., & Mattia, D. (2014). The User-Centered Design as Novel Perspective for Evaluating the Usability of BCI-Controlled Applications. *PLoS ONE*, *9*(12), Article 12. https://doi.org/10.1371/journal.pone.0112392

Kuebler, A., Kotchoubey, B., Salzmann, H. P., Ghanayim, N., Perelmouter, J., Hömberg, V., & Birbaumer, N. (1998). Self-regulation of slow cortical potentials in completely paralyzed human patients. *Neuroscience Letters*, *252*(3), Article 3.

Kukleta, M., Damborská, A., Roman, R., Rektor, I., & Brázdil, M. (2016). The primary motor cortex is involved in the control of a non-motor cognitive action. *Clinical Neurophysiology: Official*

*Journal of the International Federation of Clinical Neurophysiology*, *127*(2), Article 2. https://doi.org/10.1016/j.clinph.2015.11.049

Kumru, H., Vidal, J., Kofler, M., Benito, J., Garcia, A., & Valls-Solé, J. (2008). Exaggerated auditory startle responses in patients with spinal cord injury. *Journal of Neurology*, *255*(5), 703-709. https://doi.org/10.1007/s00415-008-0780-3

Kustov, A. A., & Robinson, D. L. (1995). Modified saccades evoked by stimulation of the macaque superior colliculus account for properties of the resettable integrator. *Journal of Neurophysiology*, *73*(4), 1724-1728. https://doi.org/10.1152/jn.1995.73.4.1724

Kyselo, M. (2020). More than our Body : Minimal and Enactive Selfhood in Global Paralysis. *Neuroethics*, *13*(2), 203-220. https://doi.org/10.1007/s12152-019-09404-9

Kyselo, M., & Di Paolo, E. (2015). Locked-in syndrome : A challenge for embodied cognitive science. *Phenomenology and the Cognitive Sciences*, *14*(3), Article 3. https://doi.org/10.1007/s11097-013-9344-9

Lakatos, P., Karmos, G., Mehta, A. D., Ulbert, I., & Schroeder, C. E. (2008). Entrainment of Neuronal Oscillations as a Mechanism of Attentional Selection. *Science*. https://doi.org/10.1126/science.1154735

Larson, E., & Lee, A. K. C. (2013). The cortical dynamics underlying effective switching of auditory spatial attention. *NeuroImage*, *64*, 365-370. https://doi.org/10.1016/j.neuroimage.2012.09.006

Lazarou, I., Nikolopoulos, S., Petrantonakis, P. C., Kompatsiaris, I., & Tsolaki, M. (2018). EEG-Based Brain–Computer Interfaces for Communication and Rehabilitation of People with Motor Impairment : A Novel Approach of the 21st Century. *Frontiers in Human Neuroscience*, *12*. https://doi.org/10.3389/fnhum.2018.00014

Lebedev, M. A. (2017). Commentary : Cortical activity in the null space: permitting preparation without movement. *Frontiers in Neuroscience*, *11*. https://doi.org/10.3389/fnins.2017.00502

Lederman, S. J., & Abbott, S. G. (1981). Texture perception : Studies of intersensory organization using a discrepancy paradigm, and visual versus tactual psychophysics. *Journal of Experimental Psychology: Human Perception and Performance*, *7*(4), 902-915. https://doi.org/10.1037/0096-1523.7.4.902

Leinders, S., Vansteensel, M. J., Branco, M. P., Freudenburg, Z. V., Pels, E. G. M., Van der Vijgh, B., Van Zandvoort, M. J. E., Ramsey, N. F., & Aarnoutse, E. J. (2020). Dorsolateral prefrontal cortex-based control with an implanted brain–computer interface. *Scientific Reports*, *10*(1), 15448. https://doi.org/10.1038/s41598-020-71774-5

Lenggenhager, B., Pazzaglia, M., Scivoletto, G., Molinari, M., & Aglioti, S. M. (2012). The Sense of the Body in Individuals with Spinal Cord Injury. *PLOS ONE*, *7*(11), e50757. https://doi.org/10.1371/journal.pone.0050757

Leonard, M., Renard, F., Harsan, L., Pottecher, J., Braun, M., Schneider, F., Froehlig, P., Blanc, F., Roquet, D., Achard, S., Meyer, N., & Kremer, S. (2019). Diffusion tensor imaging reveals diffuse white matter injuries in locked-in syndrome patients. *PLOS ONE*, *14*(4), Article 4. https://doi.org/10.1371/journal.pone.0213528

León-Carrión, J., van Eeckhout, P., Domínguez-Morales, M. D. R., & Pérez-Santamaría, F. J. (2002). The locked-in syndrome : A syndrome looking for a therapy. *Brain Injury*, *16*(7), Article 7. https://doi.org/10.1080/02699050110119781

Lesenfants, D., Habbal, D., Chatelle, C., Soddu, A., Laureys, S., & Noirhomme, Q. (2018). Toward an Attention-Based Diagnostic Tool for Patients With Locked-in Syndrome. *Clinical EEG and Neuroscience*, *49*(2), 122-135. https://doi.org/10.1177/1550059416674842

Lesenfants, D., Habbal, D., Lugo, Z., Lebeau, M., Horki, P., Amico, E., Pokorny, C., Gómez, F., Soddu, A., Müller-Putz, G., Laureys, S., & Noirhomme, Q. (2014). An independent SSVEP-based brain–computer interface in locked-in syndrome. *Journal of Neural Engineering*, *11*(3), 035002. https://doi.org/10.1088/1741-2560/11/3/035002

Li, H.-H., Hanning, N. M., & Carrasco, M. (2021). To look or not to look : Dissociating presaccadic and covert spatial attention. *Trends in Neurosciences*, *44*(8), 669-686. https://doi.org/10.1016/j.tins.2021.05.002

Li, H.-H., Pan, J., & Carrasco, M. (2021). Different computations underlie overt presaccadic and covert spatial attention. *Nature Human Behaviour*, 1-14. https://doi.org/10.1038/s41562-021-01099-4

Li, S., Chang, S.-H., Francisco, G. E., & Verduzco-Gutierrez, M. (2014). Acoustic startle reflex in patients with chronic stroke at different stages of motor recovery : A pilot study. *Topics in Stroke Rehabilitation*, *21*(4), 358-370. https://doi.org/10.1310/tsr2104-358

Libertus, K., & Hauf, P. (2017). Editorial : Motor Skills and Their Foundational Role for Perceptual, Social, and Cognitive Development. *Frontiers in Psychology*, *8*. https://www.frontiersin.org/articles/10.3389/fpsyg.2017.00301

Liik, M., Puksa, L., Lüüs, S.-M., Haldre, S., & Taba, P. (2012). Fulminant inflammatory neuropathy mimicking cerebral death. *BMJ Case Reports*, *2012*, bcr1020114906. https://doi.org/10.1136/bcr-10-2011-4906

Limanowski, J. (2021). Precision control for a flexible body representation. *Neuroscience & Biobehavioral Reviews*. https://doi.org/10.1016/j.neubiorev.2021.10.023

London, J., Thompson, M., Burger, B., Hildreth, M., & Toiviainen, P. (2019). Tapping doesn't help : Synchronized self-motion and judgments of musical tempo. *Attention, Perception, & Psychophysics*, *81*(7), 2461-2472. https://doi.org/10.3758/s13414-019-01722-7

Lowet, E., Gips, B., Roberts, M. J., De Weerd, P., Jensen, O., & van der Eerden, J. (2018). Microsaccade-rhythmic modulation of neural synchronization and coding within and across cortical areas V1 and V2. *PLoS Biology*, *16*(5), e2004132. https://doi.org/10.1371/journal.pbio.2004132

Lowet, E., Gomes, B., Srinivasan, K., Zhou, H., Schafer, R. J., & Desimone, R. (2018). Enhanced Neural Processing by Covert Attention only during Microsaccades Directed toward the Attended Stimulus. *Neuron*, *99*(1), 207-214.e3. https://doi.org/10.1016/j.neuron.2018.05.041

Lugo, Z. R., Pokorny, C., Pellas, F., Noirhomme, Q., Laureys, S., Müller-Putz, G., & Kübler, A. (2019). Mental imagery for brain-computer interface control and communication in non-responsive individuals. *Annals of Physical and Rehabilitation Medicine*. https://doi.org/10.1016/j.rehab.2019.02.005

Lugo, Z. R., Quitadamo, L. R., Bianchi, L., Pellas, F., Veser, S., Lesenfants, D., Real, R. G. L., Herbert, C., Guger, C., Kotchoubey, B., Mattia, D., Kübler, A., Laureys, S., & Noirhomme, Q. (2016).

Cognitive Processing in Non-Communicative Patients : What Can Event-Related Potentials Tell Us? *Frontiers in Human Neuroscience*, *10*, 569. https://doi.org/10.3389/fnhum.2016.00569

Lulé, D., Diekmann, V., Müller, H.-P., Kassubek, J., Ludolph, A. C., & Birbaumer, N. (2010). Neuroimaging of multimodal sensory stimulation in amyotrophic lateral sclerosis. *Journal of Neurology, Neurosurgery, and Psychiatry*, *81*(8), 899-906. https://doi.org/10.1136/jnnp.2009.192260

Lulé, D., Noirhomme, Q., Kleih, S. C., Chatelle, C., Halder, S., Demertzi, A., Bruno, M.-A., Gosseries, O., Vanhaudenhuyse, A., Schnakers, C., Thonnard, M., Soddu, A., Kübler, A., & Laureys, S. (2013). Probing command following in patients with disorders of consciousness using a brain-computer interface. *Clinical Neurophysiology: Official Journal of the International Federation of Clinical Neurophysiology*, *124*(1), Article 1. https://doi.org/10.1016/j.clinph.2012.04.030

Lundbye-Jensen, J., & Nielsen, J. B. (2008). Central nervous adaptations following 1 wk of wrist and hand immobilization. *Journal of Applied Physiology (Bethesda, Md.: 1985)*, *105*(1), 139-151. https://doi.org/10.1152/japplphysiol.00687.2007

Lunghi, M., Di Giorgio, E., Benavides-Varela, S., & Simion, F. (2020). Covert orienting of attention in 3-month-old infants : The case of biological motion. *Infant Behavior & Development*, *58*, 101422. https://doi.org/10.1016/j.infbeh.2020.101422

Maby, E., Perrin, M., Morlet, D., & et al. (2011). *Evaluation in a locked-in patient of the OpenViBE P300-speller* (pp. 272–275). Proceedings of the 5th International Brain-Computer Interface Conference.

Magliacano, A., Rosenfelder, M., Hieber, N., Bender, A., Estraneo, A., & Trojano, L. (2021). Spontaneous eye blinking as a diagnostic marker in prolonged disorders of consciousness. *Scientific Reports*, *11*(1), 22393. https://doi.org/10.1038/s41598-021-01858-3

Makin, T. R., Holmes, N. P., Brozzoli, C., & Farnè, A. (2012). Keeping the world at hand : Rapid visuomotor processing for hand-object interactions. *Experimental Brain Research*, *219*(4), 421-428. https://doi.org/10.1007/s00221-012-3089-5

Manaia, F., Teixeira, S., Velasques, B., Bittencourt, J., Salles, J. I., Arias-Carrión, O., Basile, L. F., Peressutti, C., de Carvalho, M. R., Cagy, M., Piedade, R., Ribeiro, P., & Machado, S. (2013).

Does immobilization of dependent hand promote adaptative changes in cerebral cortex? An analysis through qEEG asymmetry. *Neuroscience Letters*, *538*, 20-25. https://doi.org/10.1016/j.neulet.2012.12.030

Mannarelli, D., Pauletti, C., Locuratolo, N., Vanacore, N., Frasca, V., Trebbastoni, A., Inghilleri, M., & Fattapposta, F. (2014). Attentional processing in bulbar- and spinal-onset amyotrophic lateral sclerosis : Insights from event-related potentials. *Amyotrophic Lateral Sclerosis & Frontotemporal Degeneration*, *15*(1-2), Article 1-2. https://doi.org/10.3109/21678421.2013.787628

Manning, F. C., Harris, J., & Schutz, M. (2017). Temporal prediction abilities are mediated by motor effector and rhythmic expertise. *Experimental Brain Research*, *235*(3), 861-871. https://doi.org/10.1007/s00221-016-4845-8

Manning, F., & Schutz, M. (2013). « Moving to the beat » improves timing perception. *Psychonomic Bulletin & Review*, *20*(6), 1133-1139. https://doi.org/10.3758/s13423-013-0439-7

Maravita, A., Spence, C., & Driver, J. (2003). Multisensory integration and the body schema : Close to hand and within reach. *Current Biology: CB*, *13*(13), R531-539. https://doi.org/10.1016/s0960-9822(03)00449-4

Marchetti, M., Piccione, F., Silvoni, S., Gamberini, L., & Priftis, K. (2013). Covert Visuospatial Attention Orienting in a Brain-Computer Interface for Amyotrophic Lateral Sclerosis Patients. *Neurorehabilitation and Neural Repair*, 1545968312471903. https://doi.org/10.1177/1545968312471903

Marini, C., Morbelli, S., Cistaro, A., Campi, C., Caponnetto, C., Bauckneht, M., Bellini, A., Buschiazzo, A., Calamia, I., Beltrametti, M. C., Margotti, S., Fania, P., Poggi, I., Cabona, C., Capitanio, S., Piva, R., Calvo, A., Moglia, C., Canosa, A., … Sambuceti, G. (2018). Interplay between spinal cord and cerebral cortex metabolism in amyotrophic lateral sclerosis. *Brain*, *141*(8), Article 8. https://doi.org/10.1093/brain/awy152

Martikainen, M. H., Kaneko, K., & Hari, R. (2005). Suppressed responses to self-triggered sounds in the human auditory cortex. *Cerebral Cortex (New York, N.Y.: 1991)*, *15*(3), 299-302. https://doi.org/10.1093/cercor/bhh131

Masson, N., Andres, M., Carneiro pereira, S., Vandenberghe, A., Pesenti, M., & Vannuscorps, G. (2021). Shifting attention in visuospatial short-term memory does not require oculomotor planning : Insight from congenital gaze paralysis. *Neuropsychologia*, 107998. https://doi.org/10.1016/j.neuropsychologia.2021.107998

Masson, N., Andres, M., Pereira, S. C., Pesenti, M., & Vannuscorps, G. (2020). Exogenous covert shift of attention without the ability to plan eye movements. *Current Biology: CB*, *30*(18), R1032-R1033. https://doi.org/10.1016/j.cub.2020.07.074

McCane, L. M., Heckman, S. M., McFarland, D. J., Townsend, G., Mak, J. N., Sellers, E. W., Zeitlin, D., Tenteromano, L. M., Wolpaw, J. R., & Vaughan, T. M. (2015). P300-based brain-computer interface (BCI) event-related potentials (ERPs) : People with amyotrophic lateral sclerosis (ALS) vs. age-matched controls. *Clinical Neurophysiology: Official Journal of the International Federation of Clinical Neurophysiology*, *126*(11), Article 11. https://doi.org/10.1016/j.clinph.2015.01.013

McIsaac, T. L., Fritz, N. E., Quinn, L., & Muratori, L. M. (2018). Cognitive-Motor Interference in Neurodegenerative Disease : A Narrative Review and Implications for Clinical Management. *Frontiers in Psychology*, *9*. https://doi.org/10.3389/fpsyg.2018.02061

McKay, L. C., Evans, K. C., Frackowiak, R. S. J., & Corfield, D. R. (2003). Neural correlates of voluntary breathing in humans. *Journal of Applied Physiology*, *95*(3), 1170-1178. https://doi.org/10.1152/japplphysiol.00641.2002

McMackin, R., Dukic, S., Broderick, M., Iyer, P. M., Pinto-Grau, M., Mohr, K., Chipika, R., Coffey, A., Buxo, T., Schuster, C., Gavin, B., Heverin, M., Bede, P., Pender, N., Lalor, E. C., Muthuraman, M., Hardiman, O., & Nasseroleslami, B. (2019). Dysfunction of attention switching networks in amyotrophic lateral sclerosis. *NeuroImage : Clinical*, *22*. https://doi.org/10.1016/j.nicl.2019.101707

McPeek, R. M., & Keller, E. L. (2004). Deficits in saccade target selection after inactivation of superior colliculus. *Nature Neuroscience*, *7*(7), 757-763. https://doi.org/10.1038/nn1269

Mendoza, G., & Merchant, H. (2014). Motor system evolution and the emergence of high cognitive functions. *Progress in Neurobiology*, *122*, 73-93. https://doi.org/10.1016/j.pneurobio.2014.09.001

Merchant, H., Harrington, D. L., & Meck, W. H. (2013). Neural basis of the perception and estimation of time. *Annual Review of Neuroscience*, *36*, 313-336. https://doi.org/10.1146/annurev-neuro-062012-170349

Mirza, M. B., Adams, R. A., Friston, K., & Parr, T. (2019). Introducing a Bayesian model of selective attention based on active inference. *Scientific Reports*, *9*(1), Article 1. https://doi.org/10.1038/s41598-019-50138-8

Miyawaki, Y., Otani, T., & Morioka, S. (2020a). Agency judgments in post-stroke patients with sensorimotor deficits. *PloS One*, *15*(3), e0230603. https://doi.org/10.1371/journal.pone.0230603

Miyawaki, Y., Otani, T., & Morioka, S. (2020b). Dynamic Relationship between Sense of Agency and Post-Stroke Sensorimotor Deficits : A Longitudinal Case Study. *Brain Sciences*, *10*(5). https://doi.org/10.3390/brainsci10050294

Mizutani, T., Sakamaki, S., Tsuchiya, N., Kamei, S., Kohzu, H., Horiuchi, R., Ida, M., Shiozawa, R., & Takasu, T. (1992). Amyotrophic lateral sclerosis with ophthalmoplegia and multisystem degeneration in patients on long-term use of respirators. *Acta Neuropathologica*, *84*(4), Article 4. https://doi.org/10.1007/BF00227663

Mole, C. (2017). Attention. In E. N. Zalta (Éd.), *The Stanford Encyclopedia of Philosophy* (Fall 2017). Metaphysics Research Lab, Stanford University. https://plato.stanford.edu/archives/fall2017/entries/attention/

Monosov, I. E., Trageser, J. C., & Thompson, K. G. (2008). Measurements of Simultaneously Recorded Spiking Activity and Local Field Potentials Suggest that Spatial Selection Emerges in the Frontal Eye Field. *Neuron*, *57*(4), Article 4. https://doi.org/10.1016/j.neuron.2007.12.030

Monti, M. M., Vanhaudenhuyse, A., Coleman, M. R., Boly, M., Pickard, J. D., Tshibanda, L., Owen, A. M., & Laureys, S. (2010). Willful modulation of brain activity in disorders of consciousness.

*The New England Journal of Medicine*, *362*(7), Article 7. https://doi.org/10.1056/NEJMoa0905370

Morlet, D., Mattout, J., Fischer, C., Luauté, J., Dailler, F., Ruby, P., & André-Obadia, N. (2022). Infraclinical detection of voluntary attention in coma and post-coma patients using electrophysiology. *Clinical Neurophysiology*. https://doi.org/10.1016/j.clinph.2022.09.019

Moss, H. E., Samelson, M., Mohan, G., & Jiang, Q. L. (2016). High and Low Contrast Visual Acuity Are Not Affected in Amyotrophic Lateral Sclerosis. *PloS One*, *11*(12), e0168714. https://doi.org/10.1371/journal.pone.0168714

Mugler, E. M., Ruf, C. A., Halder, S., Bensch, M., & Kubler, A. (2010). Design and implementation of a P300-based brain-computer interface for controlling an internet browser. *IEEE Transactions on Neural Systems and Rehabilitation Engineering: A Publication of the IEEE Engineering in Medicine and Biology Society*, *18*(6), Article 6. https://doi.org/10.1109/TNSRE.2010.2068059

Münte, T. F., Tröger, M. C., Nusser, I., Wieringa, B. M., Johannes, S., Matzke, M., & Dengler, R. (1998). Alteration of early components of the visual evoked potential in amyotrophic lateral sclerosis. *Journal of Neurology*, *245*(4), 206-210. https://doi.org/10.1007/s004150050206

Naccache, L. (2018). Minimally conscious state or cortically mediated state? *Brain*, *141*(4), Article 4. https://doi.org/10.1093/brain/awx324

Nakayama, Y., Shimizu, T., Hayashi, K., Mochizuki, Y., Nagao, M., & Oyanagi, K. (2013). [Predictors the progression of communication impairment in ALS tracheostomy ventilator users]. *Rinshō shinkeigaku = Clinical neurology*, *53*(11), Article 11.

Nam, C. S., Woo, J., & Bahn, S. (2012). Severe motor disability affects functional cortical integration in the context of brain-computer interface (BCI) use. *Ergonomics*, *55*(5), Article 5. https://doi.org/10.1080/00140139.2011.647095

Neisser, U. (1976). *Cognition and Reality : Principles and Implications of Cognitive Psychology*. W. H. Freeman.

Neumann, N., & Kotchoubey, B. (2004). Assessment of cognitive functions in severely paralysed and severely brain-damaged patients : Neuropsychological and electrophysiological methods. *Brain Research Protocols*, *14*(1), Article 1. https://doi.org/10.1016/j.brainresprot.2004.09.001

Neumann, N., Kübler, A., Kaiser, J., Hinterberger, T., & Birbaumer, N. (2003). Conscious perception of brain states : Mental strategies for brain-computer communication. *Neuropsychologia*, *41*(8), Article 8.

Neumann, O. (1987). Beyond capacity : A functional view of attention. In *Perspectives on perception and action* (p. 361-394). Lawrence Erlbaum Associates, Inc.

Newbold, D. J., Gordon, E. M., Laumann, T. O., Seider, N. A., Montez, D. F., Gross, S. J., Zheng, A., Nielsen, A. N., Hoyt, C. R., Hampton, J. M., Ortega, M., Adeyemo, B., Miller, D. B., Van, A. N., Marek, S., Schlaggar, B. L., Carter, A. R., Kay, B. P., Greene, D. J., … Dosenbach, N. U. F. (2021). Cingulo-opercular control network and disused motor circuits joined in standby mode. *Proceedings of the National Academy of Sciences of the United States of America*, *118*(13), e2019128118. https://doi.org/10.1073/pnas.2019128118

Newbold, D. J., Laumann, T. O., Hoyt, C. R., Hampton, J. M., Montez, D. F., Raut, R. V., Ortega, M., Mitra, A., Nielsen, A. N., Miller, D. B., Adeyemo, B., Nguyen, A. L., Scheidter, K. M., Tanenbaum, A. B., Van, A. N., Marek, S., Schlaggar, B. L., Carter, A. R., Greene, D. J., … Dosenbach, N. U. F. (2020). Plasticity and Spontaneous Activity Pulses in Disused Human Brain Circuits. *Neuron*, *107*(3), 580-589.e6. https://doi.org/10.1016/j.neuron.2020.05.007

Nijboer, F. (2015). Technology transfer of brain-computer interfaces as assistive technology : Barriers and opportunities. *Annals of Physical and Rehabilitation Medicine*, *58*(1), Article 1. https://doi.org/10.1016/j.rehab.2014.11.001

Nobre, A. C., Gitelman, D. R., Dias, E. C., & Mesulam, M. M. (2000). Covert visual spatial orienting and saccades : Overlapping neural systems. *NeuroImage*, *11*(3), Article 3. https://doi.org/10.1006/nimg.2000.0539

Noel, J.-P., Chatelle, C., Perdikis, S., Jöhr, J., Lopes Da Silva, M., Ryvlin, P., De Lucia, M., Millán, J. del R., Diserens, K., & Serino, A. (2019). Peri-personal space encoding in patients with disorders of consciousness and cognitive-motor dissociation. *NeuroImage: Clinical*, *24*, 101940. https://doi.org/10.1016/j.nicl.2019.101940

Norman, D. A., & Shallice, T. (1986). Attention to Action. In R. J. Davidson, G. E. Schwartz, & D. Shapiro (Éds.), *Consciousness and Self-Regulation : Advances in Research and Theory Volume 4* (p. 1-18). Springer US. https://doi.org/10.1007/978-1-4757-0629-1_1

Okahara, Y., Takano, K., Nagao, M., Kondo, K., Iwadate, Y., Birbaumer, N., & Kansaku, K. (2018). Long-term use of a neural prosthesis in progressive paralysis. *Scientific Reports*, *8*(1), Article 1. https://doi.org/10.1038/s41598-018-35211-y

Orgs, G., Dovern, A., Hagura, N., Haggard, P., Fink, G. R., & Weiss, P. H. (2016). Constructing Visual Perception of Body Movement with the Motor Cortex. *Cerebral Cortex (New York, N.Y.: 1991)*, *26*(1), 440-449. https://doi.org/10.1093/cercor/bhv262

Ostarek, M., & Bottini, R. (2021). Towards Strong Inference in Research on Embodiment—Possibilities and Limitations of Causal Paradigms. *Journal of Cognition*, *4*(1), 5. https://doi.org/10.5334/joc.139

Owen, A. M., Coleman, M. R., Boly, M., Davis, M. H., Laureys, S., & Pickard, J. D. (2006). Detecting awareness in the vegetative state. *Science (New York, N.Y.)*, *313*(5792), Article 5792. https://doi.org/10.1126/science.1130197

Park, H.-D., Barnoud, C., Trang, H., Kannape, O. A., Schaller, K., & Blanke, O. (2020). Breathing is coupled with voluntary action and the cortical readiness potential. *Nature Communications*, *11*. https://doi.org/10.1038/s41467-019-13967-9

Park, H.-D., Piton, T., Kannape, O. A., Duncan, N. W., Lee, K.-Y., Lane, T. J., & Blanke, O. (2022). Breathing is coupled with voluntary initiation of mental imagery. *NeuroImage*, *264*, 119685. https://doi.org/10.1016/j.neuroimage.2022.119685

Parr, T., & Friston, K. J. (2019). Attention or salience? *Current Opinion in Psychology*, *29*, 1-5. https://doi.org/10.1016/j.copsyc.2018.10.006

Pastukhov, A., & Braun, J. (2010). Rare but precious : Microsaccades are highly informative about attentional allocation. *Vision Research*, *50*(12), 1173-1184. https://doi.org/10.1016/j.visres.2010.04.007

Pekkonen, E., Osipova, D., & Laaksovirta, H. (2004). Magnetoencephalographic evidence of abnormal auditory processing in amyotrophic lateral sclerosis with bulbar signs. *Clinical Neurophysiology*, *115*(2), Article 2. https://doi.org/10.1016/S1388-2457(03)00360-2

Perez, P., Valente, M., Hermann, B., Sitt, J., Faugeras, F., Demeret, S., Rohaut, B., & Naccache, L. (2021). Auditory Event-Related "Global Effect" Predicts Recovery of Overt Consciousness. *Frontiers in Neurology*, *11*. https://www.frontiersin.org/articles/10.3389/fneur.2020.588233

Perl, O., Ravia, A., Rubinson, M., Eisen, A., Soroka, T., Mor, N., Secundo, L., & Sobel, N. (2019). Human non-olfactory cognition phase-locked with inhalation. *Nature Human Behaviour*, *3*(5), 501-512. https://doi.org/10.1038/s41562-019-0556-z

Perrin, F., Schnakers, C., Schabus, M., Degueldre, C., Goldman, S., Brédart, S., Faymonville, M.-E., Lamy, M., Moonen, G., Luxen, A., Maquet, P., & Laureys, S. (2006). Brain response to one's own name in vegetative state, minimally conscious state, and locked-in syndrome. *Archives of Neurology*, *63*(4), Article 4. https://doi.org/10.1001/archneur.63.4.562

Pfurtscheller, G., Guger, C., Müller, G., Krausz, G., & Neuper, C. (2000). Brain oscillations control hand orthosis in a tetraplegic. *Neuroscience Letters*, *292*(3), Article 3. https://doi.org/10.1016/S0304-3940(00)01471-3

Pfurtscheller, G., Neuper, C., Müller, G. R., Obermaier, B., Krausz, G., Schlögl, A., Scherer, R., Graimann, B., Keinrath, C., Skliris, D., Wörtz, M., Supp, G., & Schrank, C. (2003). Graz-BCI : State of the art and clinical applications. *IEEE Transactions on Neural Systems and Rehabilitation Engineering: A Publication of the IEEE Engineering in Medicine and Biology Society*, *11*(2), 177-180. https://doi.org/10.1109/TNSRE.2003.814454

Pistoia, F., Cornia, R., Conson, M., Gosseries, O., Carolei, A., Sacco, S., Quattrocchi, C. C., Mallio, C. A., Iani, C., Mambro, D. D., & Sarà, M. (2016). Disembodied Mind : Cortical Changes Following Brainstem Injury in Patients with Locked-in Syndrome. *The Open Neuroimaging Journal*, *10*, 32-40. https://doi.org/10.2174/1874440001610010032

Plum, F., & Posner, J. B. (1972). The diagnosis of stupor and coma. *Contemporary Neurology Series*, *10*, 1-286.

Poiroux, E., Cavaro-Ménard, C., Leruez, S., Lemée, J. M., Richard, I., & Dinomais, M. (2015). What Do Eye Gaze Metrics Tell Us about Motor Imagery? *PLoS ONE*, *10*(11), e0143831. https://doi.org/10.1371/journal.pone.0143831

Polich, J. (2007). Updating P300 : An integrative theory of P3a and P3b. *Clinical Neurophysiology: Official Journal of the International Federation of Clinical Neurophysiology*, *118*(10), Article 10. https://doi.org/10.1016/j.clinph.2007.04.019

Pollack, I., & Rose, M. (1967). Effect of head movement on the localization of sounds in the equatorial plane. *Perception & Psychophysics*, *2*(12), 591-596. https://doi.org/10.3758/BF03210274

Pugdahl, K., Fuglsang-Frederiksen, A., de Carvalho, M., Johnsen, B., Fawcett, P. R. W., Labarre-Vila, A., Liguori, R., Nix, W. A., & Schofield, I. S. (2007). Generalised sensory system abnormalities in amyotrophic lateral sclerosis : A European multicentre study. *Journal of Neurology, Neurosurgery, and Psychiatry*, *78*(7), Article 7. https://doi.org/10.1136/jnnp.2006.098533

Puig, M. S., Zapata, L. P., Aznar-Casanova, J. A., & Supèr, H. (2013). A Role of Eye Vergence in Covert Attention. *PLOS ONE*, *8*(1), e52955. https://doi.org/10.1371/journal.pone.0052955

Ragazzoni, A., Grippo, A., Tozzi, F., & Zaccara, G. (2000). Event-related potentials in patients with total locked-in state due to fulminant Guillain–Barré syndrome. *International Journal of Psychophysiology*, *37*(1), 99-109. https://doi.org/10.1016/S0167-8760(00)00098-2

Raggi, A., Consonni, M., Iannaccone, S., Perani, D., Zamboni, M., Sferrazza, B., & Cappa, S. F. (2008). Auditory event-related potentials in non-demented patients with sporadic amyotrophic lateral sclerosis. *Clinical Neurophysiology*, *119*(2), Article 2. https://doi.org/10.1016/j.clinph.2007.10.010

Regan, M. P., & Regan, D. (1989). Objective investigation of visual function using a nondestructive zoom-FFT technique for evoked potential analysis. *The Canadian Journal of Neurological Sciences. Le Journal Canadien Des Sciences Neurologiques*, *16*(2), Article 2.

Reyes-Leiva, D., Dols-Icardo, O., Sirisi, S., Cortés-Vicente, E., Turon-Sans, J., de Luna, N., Blesa, R., Belbin, O., Montal, V., Alcolea, D., Fortea, J., Lleó, A., Rojas-García, R., & Illán-Gala, I. (2022). Pathophysiological Underpinnings of Extra-Motor Neurodegeneration in Amyotrophic

Lateral Sclerosis : New Insights From Biomarker Studies. *Frontiers in Neurology*, *12*, 750543. https://doi.org/10.3389/fneur.2021.750543

Rizzolatti, G., Riggio, L., Dascola, I., & Umiltá, C. (1987). Reorienting attention across the horizontal and vertical meridians : Evidence in favor of a premotor theory of attention. *Neuropsychologia*, *25*(1), Article 1. https://doi.org/10.1016/0028-3932(87)90041-8

Rolfs, M., Kliegl, R., & Engbert, R. (2008). Toward a model of microsaccade generation : The case of microsaccadic inhibition. *Journal of Vision*, *8*(11), 5. https://doi.org/10.1167/8.11.5

Ross, J. M., Comstock, D. C., Iversen, J. R., Makeig, S., & Balasubramaniam, R. (2022). Cortical mu rhythms during action and passive music listening. *Journal of Neurophysiology*, *127*(1), 213-224. https://doi.org/10.1152/jn.00346.2021

Rousseaux, M., Castelnot, E., Rigaux, P., Kozlowski, O., & Danzé, F. (2009). Evidence of persisting cognitive impairment in a case series of patients with locked-in syndrome. *Journal of Neurology, Neurosurgery & Psychiatry*, *80*(2), Article 2. https://doi.org/10.1136/jnnp.2007.128686

Sambo, C. F., Liang, M., Cruccu, G., & Iannetti, G. D. (2012). Defensive peripersonal space : The blink reflex evoked by hand stimulation is increased when the hand is near the face. *Journal of Neurophysiology*, *107*(3), Article 3. https://doi.org/10.1152/jn.00731.2011

Sangari, S., Iglesias, C., El Mendili, M.-M., Benali, H., Pradat, P.-F., & Marchand-Pauvert, V. (2016). Impairment of sensory-motor integration at spinal level in amyotrophic lateral sclerosis. *Clinical Neurophysiology: Official Journal of the International Federation of Clinical Neurophysiology*, *127*(4), 1968-1977. https://doi.org/10.1016/j.clinph.2016.01.014

Sato, M., Cattaneo, L., Rizzolatti, G., & Gallese, V. (2007). Numbers within Our Hands : Modulation of Corticospinal Excitability of Hand Muscles during Numerical Judgment. *Journal of Cognitive Neuroscience*, *19*(4), 684-693. https://doi.org/10.1162/jocn.2007.19.4.684

Scandola, M., Aglioti, S. M., Lazzeri, G., Avesani, R., Ionta, S., & Moro, V. (2020). Visuo-motor and interoceptive influences on peripersonal space representation following spinal cord injury. *Scientific Reports*, *10*(1), Article 1. https://doi.org/10.1038/s41598-020-62080-1

Schembri, R., Spong, J., Graco, M., Berlowitz, D. J., & COSAQ study team. (2017). Neuropsychological Function in Patients With Acute Tetraplegia and Sleep Disordered Breathing. *Sleep*, *40*(2), Article 2. https://doi.org/10.1093/sleep/zsw037

Schicatano, E. J., Mantzouranis, J., Peshori, K. R., Partin, J., & Evinger, C. (2002). Lid restraint evokes two types of motor adaptation. *The Journal of Neuroscience: The Official Journal of the Society for Neuroscience*, *22*(2), 569-576.

Schiff, N. D. (2015). Cognitive Motor Dissociation Following Severe Brain Injuries. *JAMA Neurology*, *72*(12), Article 12. https://doi.org/10.1001/jamaneurol.2015.2899

Schmidt, T. T., Jagannathan, N., Ljubljanac, M., Xavier, A., & Nierhaus, T. (2020). The multimodal Ganzfeld-induced altered state of consciousness induces decreased thalamo-cortical coupling. *Scientific Reports*, *10*(1), Article 1. https://doi.org/10.1038/s41598-020-75019-3

Schnakers, C., Bauer, C., Formisano, R., Noé, E., Llorens, R., Lejeune, N., Farisco, M., Teixeira, L., Morrissey, A.-M., De Marco, S., Veeramuthu, V., Ilina, K., Edlow, B. L., Gosseries, O., Zandalasini, M., De Bellis, F., Thibaut, A., & Estraneo, A. (2022). What names for covert awareness? A systematic review. *Frontiers in Human Neuroscience*, *16*. https://www.frontiersin.org/articles/10.3389/fnhum.2022.971315

Schnakers*, C., Majerus, S., Goldman, S., Boly*, M., Eeckhout, P. V., Gay, S., Pellas, F., Bartsch, V., Peigneux, P., Moonen, G., & Laureys*, S. (2008). Cognitive function in the locked-in syndrome. *Journal of Neurology*, *255*(3), Article 3. https://doi.org/10.1007/s00415-008-0544-0

Schnakers, C., Perrin, F., Schabus, M., Hustinx, R., Majerus, S., Moonen, G., Boly, M., Vanhaudenhuyse, A., Bruno, M.-A., & Laureys, S. (2009). Detecting consciousness in a total locked-in syndrome : An active event-related paradigm. *Neurocase*, *15*(4), Article 4. https://doi.org/10.1080/13554790902724904

Schroeder, C. E., Wilson, D. A., Radman, T., Scharfman, H., & Lakatos, P. (2010). Dynamics of Active Sensing and Perceptual Selection. *Current opinion in neurobiology*, *20*(2), 172-176. https://doi.org/10.1016/j.conb.2010.02.010

Séguin, P., Fouillen, M., Otman, A., Luauté, J., Giraux, P., Morlet, D., Maby, E., & Mattout, J. (2016). Évaluation clinique d'une interface cerveau–machine auditive à destination des personnes en

Locked-in syndrome complet. *Neurophysiologie Clinique/Clinical Neurophysiology*, *46*(2), Article 2. https://doi.org/10.1016/j.neucli.2016.05.066

Séguin, P., Maby, E., Fouillen, M., Otman, A., Luauté, J., Giraux, P., Morlet, D., & Mattout, J. (2023). *The challenge of controlling an auditory BCI in the case of severe motor disability* (p. 2023.01.10.23284295). medRxiv. https://doi.org/10.1101/2023.01.10.23284295

Sellers, E. W., & Donchin, E. (2006). A P300-based brain–computer interface : Initial tests by ALS patients. *Clinical Neurophysiology*, *117*(3), Article 3. https://doi.org/10.1016/j.clinph.2005.06.027

Sergent, C., Corazzol, M., Labouret, G., Stockart, F., Wexler, M., King, J.-R., Meyniel, F., & Pressnitzer, D. (2021). Bifurcation in brain dynamics reveals a signature of conscious processing independent of report. *Nature Communications*, *12*(1), Article 1. https://doi.org/10.1038/s41467-021-21393-z

Severens, M., Van der Waal, M., Farquhar, J., & Desain, P. (2014). Comparing tactile and visual gaze-independent brain-computer interfaces in patients with amyotrophic lateral sclerosis and healthy users. *Clinical Neurophysiology: Official Journal of the International Federation of Clinical Neurophysiology*. https://doi.org/10.1016/j.clinph.2014.03.005

Sherman, S. M., & Usrey, W. M. (2021). Cortical control of behavior and attention from an evolutionary perspective. *Neuron*, *109*(19), 3048-3054. https://doi.org/10.1016/j.neuron.2021.06.021

Simões-Franklin, C., Whitaker, T. A., & Newell, F. N. (2011). Active and passive touch differentially activate somatosensory cortex in texture perception. *Human Brain Mapping*, *32*(7), 1067-1080. https://doi.org/10.1002/hbm.21091

Smith, D. T., Rorden, C., & Jackson, S. R. (2004). Exogenous orienting of attention depends upon the ability to execute eye movements. *Current Biology: CB*, *14*(9), 792-795. https://doi.org/10.1016/j.cub.2004.04.035

Smith, D. T., & Schenk, T. (2012). The Premotor theory of attention : Time to move on? *Neuropsychologia*, *50*(6), Article 6. https://doi.org/10.1016/j.neuropsychologia.2012.01.025

Smith, E., & Delargy, M. (2005). Locked-in syndrome. *BMJ : British Medical Journal*, *330*(7488), Article 7488.

Smith, P. F. (2019). The Growing Evidence for the Importance of the Otoliths in Spatial Memory. *Frontiers in Neural Circuits*, *13*, 66. https://doi.org/10.3389/fncir.2019.00066

Sorrentino, P., Rucco, R., Jacini, F., Trojsi, F., Lardone, A., Baselice, F., Femiano, C., Santangelo, G., Granata, C., Vettoliere, A., Monsurrò, M. R., Tedeschi, G., & Sorrentino, G. (2018). Brain functional networks become more connected as amyotrophic lateral sclerosis progresses : A source level magnetoencephalographic study. *NeuroImage. Clinical*, *20*, 564-571. https://doi.org/10.1016/j.nicl.2018.08.001

Spueler, M. (2018). No Evidence for Communication in the Complete Locked-in State. *bioRxiv*, 287631. https://doi.org/10.1101/287631

Stevens, J. K., Emerson, R. C., Gerstein, G. L., Kallos, T., Neufeld, G. R., Nichols, C. W., & Rosenquist, A. C. (1976). Paralysis of the awake human : Visual perceptions. *Vision Research*, *16*(1), 93-IN9. https://doi.org/10.1016/0042-6989(76)90082-1

Stoll, J., Chatelle, C., Carter, O., Koch, C., Laureys, S., & Einhäuser, W. (2013). Pupil responses allow communication in locked-in syndrome patients. *Current Biology: CB*, *23*(15), R647-648. https://doi.org/10.1016/j.cub.2013.06.011

Strauch, C., Wang, C.-A., Einhäuser, W., Van der Stigchel, S., & Naber, M. (2022). Pupillometry as an integrated readout of distinct attentional networks. *Trends in Neurosciences*. https://doi.org/10.1016/j.tins.2022.05.003

Strauss, D. J., Corona-Strauss, F. I., Schroeer, A., Flotho, P., Hannemann, R., & Hackley, S. A. (2020). Vestigial auriculomotor activity indicates the direction of auditory attention in humans. *eLife*, *9*, e54536. https://doi.org/10.7554/eLife.54536

Taylor, J. P., Brown, R. H., & Cleveland, D. W. (2016). Decoding ALS : From genes to mechanism. *Nature*, *539*(7628), Article 7628. https://doi.org/10.1038/nature20413

Thompson, K. G. (2005). Neuronal Basis of Covert Spatial Attention in the Frontal Eye Field. *Journal of Neuroscience*, *25*(41), Article 41. https://doi.org/10.1523/JNEUROSCI.0741-05.2005

Thompson, K. G., Bichot, N. P., & Schall, J. D. (1997). Dissociation of Visual Discrimination From Saccade Programming in Macaque Frontal Eye Field. *Journal of Neurophysiology*, *77*(2), 1046-1050. https://doi.org/10.1152/jn.1997.77.2.1046

Tolonen, M., Palva, J. M., Andersson, S., & Vanhatalo, S. (2007). Development of the spontaneous activity transients and ongoing cortical activity in human preterm babies. *Neuroscience*, *145*(3), 997-1006. https://doi.org/10.1016/j.neuroscience.2006.12.070

Treder, M. S., & Blankertz, B. (2010). (C)overt attention and visual speller design in an ERP-based brain-computer interface. *Behavioral and Brain Functions*, *6*(1), Article 1. https://doi.org/10.1186/1744-9081-6-28

Troxler, D. (1804). *Uber das Verschwindern gegebener Gegenstande innerhalb unsers Gesichtskreises.*

Tsakiris, M., Tajadura-Jiménez, A., & Costantini, M. (2011). Just a heartbeat away from one's body : Interoceptive sensitivity predicts malleability of body-representations. *Proceedings. Biological Sciences*, *278*(1717), 2470-2476. https://doi.org/10.1098/rspb.2010.2547

Turano, K. (1990). Bisection Accuracy and Precision in Patients with Retinitis Pigmentosa. *Noninvasive Assessment of the Visual System (1990), Paper WA2*, WA2. https://doi.org/10.1364/NAVS.1990.WA2

Turano, K. A., Yu, D., Hao, L., & Hicks, J. C. (2005). Optic-flow and egocentric-direction strategies in walking : Central vs peripheral visual field. *Vision Research*, *45*(25-26), 3117-3132. https://doi.org/10.1016/j.visres.2005.06.017

Valsecchi, M., Betta, E., & Turatto, M. (2007). Visual oddballs induce prolonged microsaccadic inhibition. *Experimental Brain Research*, *177*(2), 196-208. https://doi.org/10.1007/s00221-006-0665-6

Valsecchi, M., & Turatto, M. (2009). Microsaccadic responses in a bimodal oddball task. *Psychological Research*, *73*(1), 23-33. https://doi.org/10.1007/s00426-008-0142-x

van Beers, R. J., Wolpert, D. M., & Haggard, P. (2002). When Feeling Is More Important Than Seeing in Sensorimotor Adaptation. *Current Biology*, *12*(10), 834-837. https://doi.org/10.1016/S0960-9822(02)00836-9

van Moorselaar, D., Lampers, E., Cordesius, E., & Slagter, H. A. (2020). Neural mechanisms underlying expectation-dependent inhibition of distracting information. *eLife*, *9*, e61048. https://doi.org/10.7554/eLife.61048

Vansteensel, M. J., Bleichner, M. G., Freudenburg, Z. V., Hermes, D., Aarnoutse, E. J., Leijten, F. S. S., Ferrier, C. H., Jansma, J. M., & Ramsey, N. F. (2014). Spatiotemporal characteristics of electrocortical brain activity during mental calculation. *Human Brain Mapping*, *35*(12), 5903-5920. https://doi.org/10.1002/hbm.22593

Vb, P., & Rj, Z. (2019, mars 6). *Rhythm and time in the premotor cortex*. PLoS Biology; PLoS Biol. https://doi.org/10.1371/journal.pbio.3000293

Vecera, S. P., Cosman, J. D., Vatterott, D. B., & Roper, Z. J. J. (2014). Chapter Eight - The Control of Visual Attention : Toward a Unified Account. In B. H. Ross (Éd.), *Psychology of Learning and Motivation* (Vol. 60, p. 303-347). Academic Press. https://doi.org/10.1016/B978-0-12-800090-8.00008-1

Verhagen, W. I., Huygen, P. L., & Schulte, B. P. (1986). Clinical and electrophysiological study in a patient surviving from locked-in syndrome. *Clinical Neurology and Neurosurgery*, *88*(1), Article 1.

Verleger, R. (2020). Effects of relevance and response frequency on P3b amplitudes : Review of findings and comparison of hypotheses about the process reflected by P3b. *Psychophysiology*, *57*(7), e13542. https://doi.org/10.1111/psyp.13542

Verleger, R., Hamann, L. M., Asanowicz, D., & Śmigasiewicz, K. (2015). Testing the S-R link hypothesis of P3b : The oddball effect on S1-evoked P3 gets reduced by increased task relevance of S2. *Biological Psychology*, *108*, 25-35. https://doi.org/10.1016/j.biopsycho.2015.02.010

Vernet, M., Quentin, R., Chanes, L., Mitsumasu, A., & Valero-Cabré, A. (2014). Frontal eye field, where art thou? Anatomy, function, and non-invasive manipulation of frontal regions involved in eye movements and associated cognitive operations. *Frontiers in Integrative Neuroscience*, *8*. https://doi.org/10.3389/fnint.2014.00066

Verrillo, R. T., Bolanowski, S. J., & McGlone, F. P. (1999). Subjective magnitude of tactile roughness. *Somatosensory & Motor Research*, *16*(4), 352-360. https://doi.org/10.1080/08990229970401

Vidaurre, C., & Blankertz, B. (2010). Towards a cure for BCI illiteracy. *Brain Topography*, *23*(2), Article 2. https://doi.org/10.1007/s10548-009-0121-6

Vieregge, P., Wauschkuhn, B., Heberlein, I., Hagenah, J., & Verleger, R. (1999). Selective attention is impaired in amyotrophic lateral sclerosis—A study of event-related EEG potentials. *Brain Research. Cognitive Brain Research*, *8*(1), 27-35. https://doi.org/10.1016/s0926-6410(99)00004-x

Vilares, I., & Kording, K. (2011). Bayesian models : The structure of the world, uncertainty, behavior, and the brain. *Annals of the New York Academy of Sciences*, *1224*, 22-39. https://doi.org/10.1111/j.1749-6632.2011.05965.x

Vinck, M., Batista-Brito, R., Knoblich, U., & Cardin, J. A. (2015). Arousal and locomotion make distinct contributions to cortical activity patterns and visual encoding. *Neuron*, *86*(3), 740-754. https://doi.org/10.1016/j.neuron.2015.03.028

Volpato, C., Prats Sedano, M. A., Silvoni, S., Segato, N., Cavinato, M., Merico, A., Piccione, F., Palmieri, A., & Birbaumer, N. (2016). Selective attention impairment in amyotrophic lateral sclerosis. *Amyotrophic Lateral Sclerosis & Frontotemporal Degeneration*, *17*(3-4), 236-244. https://doi.org/10.3109/21678421.2016.1143514

Voss, M., Ingram, J. N., Wolpert, D. M., & Haggard, P. (2008). Mere expectation to move causes attenuation of sensory signals. *PloS One*, *3*(8), e2866. https://doi.org/10.1371/journal.pone.0002866

Walker, R., Techawachirakul, P., & Haggard, P. (2009). Frontal eye field stimulation modulates the balance of salience between target and distractors. *Brain Research*, *1270*, 54-63. https://doi.org/10.1016/j.brainres.2009.02.081

Walsh, E., & Haggard, P. (2007). The internal structure of stopping as revealed by a sensory detection task. *Experimental Brain Research*, *183*(3), 405-410. https://doi.org/10.1007/s00221-007-1128-4

Walsh, V. (2003). A theory of magnitude : Common cortical metrics of time, space and quantity. *Trends in Cognitive Sciences*, *7*(11), 483-488. https://doi.org/10.1016/j.tics.2003.09.002

Walter, S. (2010). Locked-in Syndrome, BCI, and a Confusion about Embodied, Embedded, Extended, and Enacted Cognition. *Neuroethics*, *3*(1), Article 1. https://doi.org/10.1007/s12152-009-9050-z

Whitham, E. M., Lewis, T., Pope, K. J., Fitzgibbon, S. P., Clark, C. R., Loveless, S., DeLosAngeles, D., Wallace, A. K., Broberg, M., & Willoughby, J. O. (2008). Thinking activates EMG in scalp electrical recordings. *Clinical Neurophysiology: Official Journal of the International Federation of Clinical Neurophysiology*, *119*(5), 1166-1175. https://doi.org/10.1016/j.clinph.2008.01.024

Whitham, E. M., Pope, K. J., Fitzgibbon, S. P., Lewis, T., Clark, C. R., Loveless, S., Broberg, M., Wallace, A., DeLosAngeles, D., Lillie, P., Hardy, A., Fronsko, R., Pulbrook, A., & Willoughby, J. O. (2007). Scalp electrical recording during paralysis : Quantitative evidence that EEG frequencies above 20 Hz are contaminated by EMG. *Clinical Neurophysiology: Official Journal of the International Federation of Clinical Neurophysiology*, *118*(8), 1877-1888. https://doi.org/10.1016/j.clinph.2007.04.027

Widmann, A., Engbert, R., & Schröger, E. (2014). Microsaccadic responses indicate fast categorization of sounds : A novel approach to study auditory cognition. *The Journal of Neuroscience: The Official Journal of the Society for Neuroscience*, *34*(33), 11152-11158. https://doi.org/10.1523/JNEUROSCI.1568-14.2014

Wijngaarden, J. B. G. van, Zucca, R., Finnigan, S., & Verschure, P. F. M. J. (2016). The Impact of Cortical Lesions on Thalamo-Cortical Network Dynamics after Acute Ischaemic Stroke : A Combined Experimental and Theoretical Study. *PLOS Computational Biology*, *12*(8), e1005048. https://doi.org/10.1371/journal.pcbi.1005048

Wolpaw, J. R., Bedlack, R. S., Reda, D. J., Ringer, R. J., Banks, P. G., Vaughan, T. M., Heckman, S. M., McCane, L. M., Carmack, C. S., Winden, S., McFarland, D. J., Sellers, E. W., Shi, H., Paine, T., Higgins, D. S., Lo, A. C., Patwa, H. S., Hill, K. J., Huang, G. D., & Ruff, R. L. (2018). Independent home use of a brain-computer interface by people with amyotrophic lateral sclerosis. *Neurology*, *91*(3), Article 3. https://doi.org/10.1212/WNL.0000000000005812

Wu, C.-T., Weissman, D. H., Roberts, K. C., & Woldorff, M. G. (2007). The neural circuitry underlying the executive control of auditory spatial attention. *Brain Research*, *1134*(1), Article 1. https://doi.org/10.1016/j.brainres.2006.11.088

Wu, W. (2019). Action always involves attention. *Analysis*, *79*(4), 693-703. https://doi.org/10.1093/analys/any080

Yamamoto, N., & Philbeck, J. W. (2013). Peripheral vision benefits spatial learning by guiding eye movements. *Memory & Cognition*, *41*(1), 109-121. https://doi.org/10.3758/s13421-012-0240-2

Yang, S. C.-H., Wolpert, D. M., & Lengyel, M. (2016). Theoretical perspectives on active sensing. *Current Opinion in Behavioral Sciences*, *11*, 100-108. https://doi.org/10.1016/j.cobeha.2016.06.009

Yin, S., Liu, Y., & Ding, M. (2016). Amplitude of Sensorimotor Mu Rhythm Is Correlated with BOLD from Multiple Brain Regions : A Simultaneous EEG-fMRI Study. *Frontiers in Human Neuroscience*, *10*. https://www.frontiersin.org/articles/10.3389/fnhum.2016.00364

Yogev-Seligmann, G., Hausdorff, J. M., & Giladi, N. (2008). The role of executive function and attention in gait. *Movement Disorders: Official Journal of the Movement Disorder Society*, *23*(3), Article 3. https://doi.org/10.1002/mds.21720

Yokosaka, T., Kuroki, S., Nishida, S., & Watanabe, J. (2015). Apparent Time Interval of Visual Stimuli Is Compressed during Fast Hand Movement. *PLoS ONE*, *10*(4), e0124901. https://doi.org/10.1371/journal.pone.0124901

Yon, D., Edey, R., Ivry, R. B., & Press, C. (2017). Time on your hands : Perceived duration of sensory events is biased toward concurrent actions. *Journal of Experimental Psychology. General*, *146*(2), 182-193. https://doi.org/10.1037/xge0000254

Yu, G., Herman, J. P., Katz, L. N., & Krauzlis, R. J. (2022). Microsaccades as a marker not a cause for attention-related modulation. *eLife*, *11*, e74168. https://doi.org/10.7554/eLife.74168

Yuval-Greenberg, S., Merriam, E. P., & Heeger, D. J. (2014). Spontaneous Microsaccades Reflect Shifts in Covert Attention. *Journal of Neuroscience*, *34*(41), 13693-13700. https://doi.org/10.1523/JNEUROSCI.0582-14.2014

Zamariola, G., Cardini, F., Mian, E., Serino, A., & Tsakiris, M. (2017). Can you feel the body that you see? On the relationship between interoceptive accuracy and body image. *Body Image*, *20*, 130-136. https://doi.org/10.1016/j.bodyim.2017.01.005

Zanini, A., Salemme, R., Farnè, A., & Brozzoli, C. (2021). Associative learning in peripersonal space : Fear responses are acquired in hand-centered coordinates. *Journal of Neurophysiology*, *126*(3), 864-874. https://doi.org/10.1152/jn.00157.2021